\documentclass[12pt]{article}
\usepackage[a4paper,margin=1in]{geometry}
\usepackage{amsmath}
\usepackage{graphicx}
\usepackage{natbib}
\usepackage{url} % not crucial - just used below for the URL 
\usepackage{changepage}
\usepackage[utf8x]{inputenc}
\usepackage{textcomp,marvosym}
\usepackage{fixltx2e}
\usepackage{amsmath,amssymb}
\usepackage{nameref,hyperref}
\usepackage{authblk}
\usepackage{algorithm}
\usepackage{algpseudocode}
\usepackage{dsfont}
\usepackage{xcolor}
\usepackage{enumitem}
\usepackage{caption,setspace}
\usepackage{pdflscape}
\usepackage{graphicx}
\usepackage{placeins}
\usepackage{mathtools}
\usepackage{appendix}
\DeclarePairedDelimiter\ceil{\lceil}{\rceil}
\DeclarePairedDelimiter\floor{\lfloor}{\rfloor}
\newcommand{\dydx}[2]{\frac{\text{d} #1}{\text{d} #2}}
\newcommand{\ddydx}[2]{\frac{\text{d}^2 #1}{\text{d} {#2}^2}}
\newcommand{\pdydx}[2]{\frac{\partial #1}{\partial #2}}
\newcommand{\pddydx}[2]{\frac{\partial^2 #1}{\partial {#2}^2}}
\newcommand{\pmddydx}[3]{\frac{\partial^2 #1}{\partial {#2}\partial {#3}}}

\renewcommand{\eqref}[1]{Equation~(\ref{#1})}
\newcommand{\Prob}[1]{\mathbb{P}(#1)}
\newcommand{\CondProb}[2]{\mathbb{P}(#1 \mid #2)}

\newcommand{\PDF}[1]{p(#1)}
\newcommand{\CondPDF}[2]{p(#1 \mid #2)}

\newcommand{\approxCondPDF}[2]{\tilde{p}(#1 \mid #2)}

\newcommand{\Kernel}[2]{q(#1\mid #2)}
\newcommand{\BKernel}[3]{Q_{#3}(#1 \mid #2)}
\newcommand{\approxBKernel}[3]{\tilde{Q}_{#3}(#1 \mid #2)}

\newcommand{\Kernelsub}[3]{q_{#3}(#1\mid #2)}
\newcommand{\approxKernelsub}[3]{\tilde{q}_{#3}(#1 \mid #2)}
\newcommand{\E}[1]{\mathbb{E}\left[#1\right]}

\newcommand{\bvec}[1]{\mathbf{#1}}
\newcommand{\ind}[2]{\mathds{1}_{#1}\left(#2\right)}

\newcommand{\simProc}[2]{s(#1 \mid #2)}
\newcommand{\approxsimProc}[2]{\tilde{s}(#1 \mid #2)}
\newcommand{\paramvec}{\boldsymbol{\theta}}
\newcommand{\approxparamvec}{\tilde{\boldsymbol{\theta}}}

\newcommand{\paramspace}{\boldsymbol{\Theta}}
\newcommand{\discrep}[2]{\rho( #1, #2)}
\newcommand{\dat}{\mathcal{D}}
\newcommand{\simdat}{\mathcal{D}_s}
\newcommand{\approxsimdat}{\tilde{\mathcal{D}}_s}

\begin{document}

\title{Rapid Bayesian inference for expensive stochastic models}

%\runtitle{A Sample Document}
%\thankstext{T1}{Footnote to the title with the ``thankstext'' command.}

\author[1,2]{David~J. Warne\footnote{To whom correspondence should be addressed. E-mail: david.warne@qut.edu.au}}
\author[3]{Ruth E. Baker}
\author[1,2]{Matthew J. Simpson}

\affil[1]{School of Mathematical Sciences, Queensland University of Technology, Brisbane, Queensland 4001, Australia}
\affil[2]{Centre for Data Science, Queensland University of Technology, Brisbane, Queensland 4001, Australia}
\affil[3]{ Mathematical Institute, University of Oxford, Oxford, OX2 6GG, United Kingdom}

\maketitle

\begin{abstract}
	Almost all fields of science rely upon statistical inference to estimate unknown parameters in theoretical and computational models. While the performance of modern computer hardware continues to grow, the computational requirements for the simulation of models are growing even faster. This is largely due to the increase in model complexity, often including stochastic dynamics, that is necessary to describe and characterize phenomena observed using modern, high resolution, experimental techniques. Such models are rarely analytically tractable, meaning that extremely large numbers of stochastic simulations are required for parameter inference. In such cases, parameter inference can be practically impossible. In this work, we present new computational Bayesian techniques that accelerate inference for expensive stochastic models by using computationally inexpensive approximations to inform feasible regions in parameter space, and through learning transforms that adjust the biased approximate inferences to closer represent the correct inferences under the expensive stochastic model. Using topical examples from ecology and cell biology, we demonstrate a speed improvement of an order of magnitude without any loss in accuracy. This represents a substantial improvement over current state-of-the-art methods for Bayesian computations when appropriate model approximations are available.
\end{abstract}


\section{Introduction}
\label{sec:intro}
%moving fronts, plants and cells
%\subsection{Growth and dispersion in biology and ecology}

%\todo{More general introduction about approximations to expensive stochastic models. ever increasing level of experimental detail, motivation to use complex stochastic models. However, connecting data through calibration is a non-trivial and unresolved problem... One can approximate the stochastic model, however, it is unclear how to capture the bias... lots of recent work connecting models of different resolutions ... here we consider the case where one has a family of models... etc ...}

Modern experimental techniques allow us to observe the natural world in unprecedented detail and resolution~\citep{Chen2014}. Advances in machine learning and artificial intelligence provide many new techniques for pattern recognition and prediction, however, in almost all scientific inquiry there is a need for detailed mathematical models to provide mechanistic insight into the phenomena observed~\citep{Baker2018,Coveney2016}. This is particularly true in the biological and ecological sciences, where detailed stochastic models are routinely applied to develop and validate theory as well as interpret and analyze data~\citep{Black2012,Drawert2017,Wilkinson2009}.

Two distinct computational challenges arise when stochastic models are considered, they are: (i) the \emph{forwards problem}; and (ii) the \emph{inverse problem}, sometimes called the \emph{backwards problem}~\citep{Warne2019b}. While the computational generation of a single sample path, that is the \emph{forwards problem}, may be feasible, generating hundreds or thousands or more such sample paths may be required to gain insight into the range of possible model predictions and to conduct parameter sensitivity analysis~\citep{Gunawan2005,Lester2017,Marino2008}. The problem is further compounded if the models must be calibrated using experimental data, that is the \emph{inverse problem} of parameter estimation, since millions of sample paths may be necessary. 

In many cases, the forwards problem can be sufficiently computationally expensive to render both parameter sensitivity analysis and the inverse problem completely intractable, despite recent advances in computational inference~\citep{Sisson2018}. This has prompted recent interest in the use of mathematical approximations to circumvent the computational burden, both in the context of the forwards and inverse problems. For example, linear approximations are applied to the forwards problem of chemical reaction networks with bimolecular and higher-order reactions~\citep{Cao2018}, and various approximations, including surrogate models~\citep{Rynn2019}, emulators~\citep{Buzbas2015} and transport maps~\citep{Parno2018}, are applied to inverse problems with expensive forwards models, for example, in the study of climate science~\citep{Holden2018}. Furthermore,
a number of developments, such as multilevel Monte Carlo methods~\citep{Giles2015}, have demonstrated that families of approximations can be combined to improve computational performance without sacrificing accuracy.  

In recent years, the Bayesian approach to the inverse problem of model calibration and parameter inference has been particularly successful in many fields of science including, astronomy~\citep{EHT2019}, anthropology and archaeology~\citep{King2014,Malaspinas2016}, paleontology and evolution~\citep{ODea2016,Pritchard1999,Tavare1997}, epidemiology~\citep{Liu2018}, biology~\citep{Lawson2018,Guindani2014,Woods2016,Vo2015}, and ecology~\citep{Ellison2004,Stumpf2014}. 
For complex stochastic models, parameterized by $\paramvec \in \paramspace$, computing the likelihood of observing data $\dat \in \mathbb{D}$ is almost always impossible~\citep{Browning2018,Johnston2014,Vankov2019}. Thus, approximate Bayesian computation (ABC) methods~\citep{Sisson2018} are essential.  ABC methods replace likelihood evaluation with an approximation based on stochastic simulations of the proposed model, this is captured directly in \emph{ABC rejection sampling}~\citep{Beaumont2002,Pritchard1999,Tavare1997} (Section~\ref{sec:methods}) where samples are generated from an approximate posterior using stochastic simulations of the forwards problem as a replacement for the likelihood. 

Unfortunately, ABC rejection sampling can be computationally expensive or even completely prohibitive, especially for high-dimensional parameter spaces, since a very large number of stochastic simulations are required to generate enough samples from the approximate Bayesian posterior distribution~\citep{Sisson2018,Warne2020}. This is further compounded when the forwards problem is computationally expensive. 
In contrast, an appropriately chosen approximate model may yield a tractable likelihood that removes the need for ABC methods~\citep{Browning2019,Warne2017a,Warne2019}. This highlights a key advantage of such approximations because no ABC sampling is required. However, approximations can perform poorly in terms of their predictive capability, and inference based on such models will always be biased, with the extent of the bias dependent on the level of accuracy.
%Thus approximations may fail to reflect the parameter uncertainty  to stochastic fluctuations in the accurate stochastic model. 

%\todo{Describe the task of Bayesian parameter inference~\citep{Gelman2014},  Highlight the intractable likelihood and hence the need for ABC (describe basic rejection scheme) methods~\citep{Beaumont2002,Pritchard1999,Tavare1997}, and the computational challenges of needing many stochastic simulations of the discrete models. However, continuum limits often have tractable likelihoods~\citep{Browning2019,Warne2017,Warne2019}, unfortunately these are not accurate in all parameter regimes~\citep{Callaghan2006}, nor do they capture additional uncertainty from model stochasticity.}

%\subsection{Contribution}

We consider ABC-based inference algorithms for the challenging problem of parameter inference for computationally expensive stochastic models when an appropriate approximation is available to inform the search in parameter space. Under our approach, the approximate model need not be quantitatively accurate in terms of the forwards problem, but must qualitatively respond to changes in parameter values in a similar way to the stochastic model. In particular, we extend the sequential Monte Carlo ABC sampler (SMC-ABC) of \citet{Sisson2007} (Section~\ref{sec:methods}) to exploit the approximate model in two ways: (i) to generate an intermediate proposal distribution, that we call a \emph{preconditioner}, to improve ABC acceptance rates for the stochastic model; and
(ii) to construct a biased ABC posterior, then reduce this bias using a \emph{moment-matching} transform.
We describe both methods and then present relevant examples from ecology and cell biology. Example calculations demonstrate that our methods generate ABC posteriors with a significant reduction in the number of required expensive stochastic simulations, leading to as much as a tenfold computational speedup. The methods we demonstrate here enable substantial acceleration of accurate statistical inference for a broad range of applications, since many areas of science utilise model approximations out of necessity despite potential inference inaccuracies. %\todo{upsell the speed improvements}

As a motivating case study for this work, we focus on stochastic models that can replicate many spatiotemporal patterns that naturally arise in biological and ecological systems. Stochastic discrete random walk models (Section~\ref{sec:results}), henceforth called \emph{discrete models}, can accurately characterize the microscale interactions of individual agents, such as animals, plants, micro-organisms, and cells~\citep{Agnew2014,Codling2008,Hardenberg2001,Law2003,Taylor2005,Vincenot2016}.  
Mathematical modeling of populations as complex systems of agents can enhance our understanding of real biological and ecological populations with applications in cancer treatment~\citep{Bottger2015}, wound healing~\citep{Callaghan2006}, wildlife conservation~\citep{McLane2011,DeAngelis2014}, and the management of invasive species~\citep{Chkrebtii2015,Taylor2005}.

\begin{landscape}
	\begin{figure}[h]
		\centering
		\includegraphics[width=\linewidth]{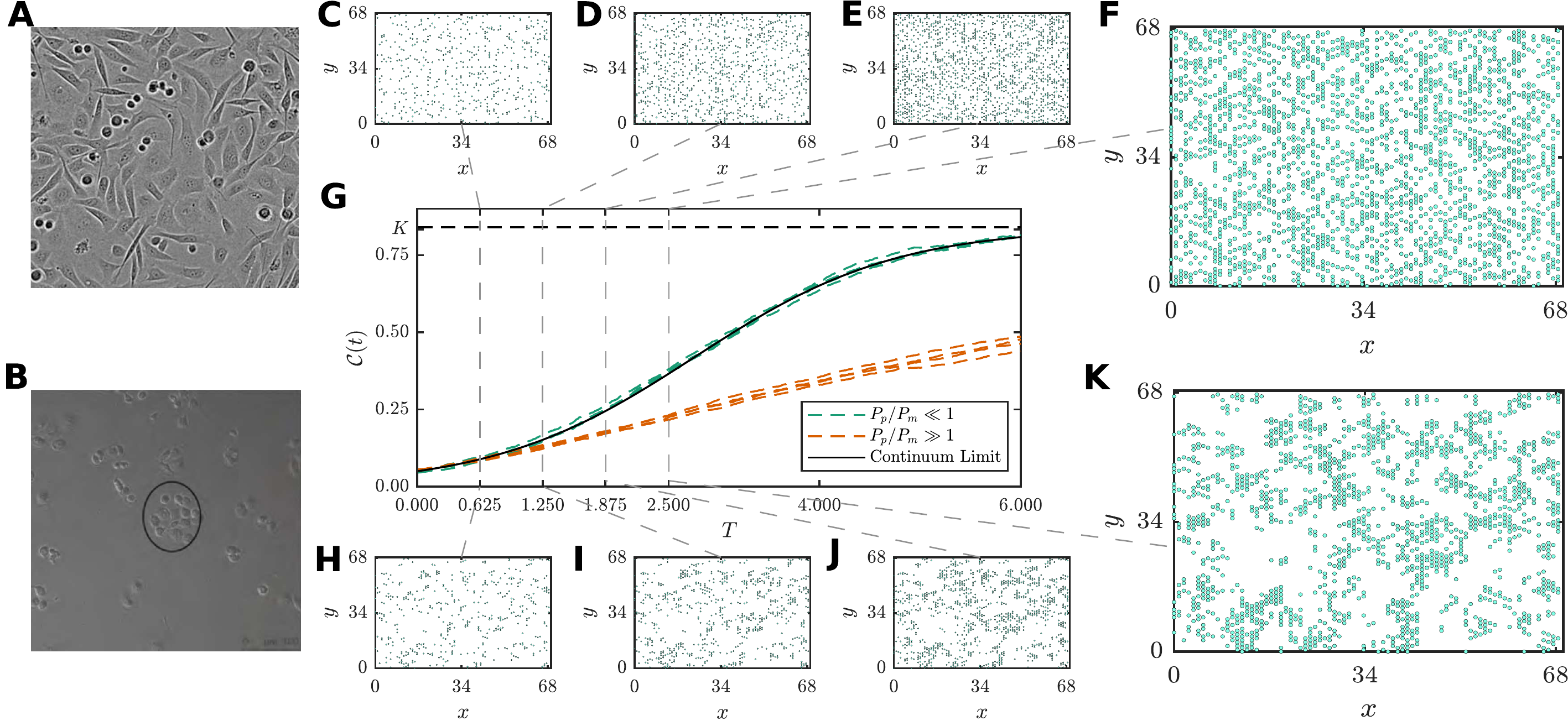}
		\caption{Discrete random walk models can replicate observed spatial patterns in cell culture: (A) PC-3 prostate cancer cells (reprinted from~\citet{Jin2017} with permission); and (B) MBA-MD-231 breast cancer cells (reprinted from~\citet{Simpson2013} with permission). (C)--(F) Discrete simulations with $P_p/P_m \ll 1$ replicate the  uniform distribution of (A) PC-3 cells. (H)--(K) Discrete simulations with $P_p/P_m \gg 1$ replicate spatial clustering of (B) MBA-MD-231 cells.  
			(G) Averaged population density profiles $\mathcal{C}(t)$ for the discrete model with highly motile agents, $P_m = 1$ (dashed green), and near stationary agents, $P_m = 5 \times 10^{-4}$ (dashed orange), compared with the logistic growth continuum limit (solid black), time is non-dimensionalized with $T = P_p t / \tau$.  
			%In all cases, discrete simulations (Section~\ref{sec:results}) are initialised with a uniformly distributed initial occupancy,  $f(C) = (1- C/K)$, $K = 5/6$ (black dashed), and $P_p = 10^{-3}$. An $I \times J$ hexagonal lattice is used with $I = 68$, and $J = 80$.
		}
		\label{fig:fig2}
	\end{figure}
\end{landscape}

For example, the discrete model formulation can replicate many realistic spatiotemporal patterns observed in cell biology. Figure~\ref{fig:fig2}(A),(B) demonstrates typical microscopy images obtained from \emph{in vitro} cell culture assays; ubiquitous and important experimental techniques used in the study of cell motility, cell proliferation and drug design. Various patterns are observed: prostate cancer cells (PC-3 line) tend to be highly motile, and spread uniformly to invade vacant spaces (Figure~\ref{fig:fig2}(A)); in contrast breast cancer cells (MBA-MD-231 line) tend be relatively stationary with proliferation events driving the formation of aggregations (Figure~\ref{fig:fig2}(B)). These phenomena may be captured using a lattice-based discrete model framework by varying the ratio $P_p/P_m$ where $P_p \in [0,1]$ and $P_m \in [0,1]$ are, respectively, the probabilities that an agent attempts to proliferate and attempts to move during a time interval of duration $\tau > 0$ (See Section \ref{sec:discrete}). For $P_p/P_m \ll 1$, behavior akin to PC-3 cells is recovered (Figure~\ref{fig:fig2}(C)--(F))~\citep{Jin2016b}. Setting $P_p/P_m \gg 1$, as in Figure~\ref{fig:fig2}(H)--(K), leads to clusters of occupied lattices sites that are similar to the aggregates of MBA-MD-231 cells~\citep{Agnew2014,Simpson2013}.

It is common practice to derive approximate continuum-limit differential equation descriptions of discrete models~\citep{Callaghan2006,Jin2016b,Simpson2010} (Supplementary Material). Such approximations provide a means of performing analysis with significantly reduced computational requirements, since evaluating an exact analytical solution, if available, or otherwise numerically solving a differential equation is typically several orders of magnitude faster than generating a single realization of the discrete model, of which hundreds or thousands may be required for reliable ABC sampling~\citep{Browning2018}. However, such approximations are generally only valid within certain parameter regimes, for example here when $P_p/P_m \ll 1$~\citep{Callaghan2006,Simpson2010}. Consider Figure~\ref{fig:fig2}(G), the population density growth curve from the continuum-limit logistic growth model is superimposed with stochastic data for four realizations of a discrete model with $P_p/P_m \ll 1$ and $P_p/P_m \gg 1$, under initial conditions simulating a proliferation assay, where each lattice site is randomly populated with constant probability, such that there are no macroscopic gradients present at $t = 0$. The continuum-limit logistic growth model is an excellent match for the $P_p/P_m \ll 1$ case (Figure~\ref{fig:fig2}(C)--(F)), but severely overestimates the population density when $P_p/P_m \gg 1$ since the mean-field assumptions underpinning the continuum-limit model are violated by the presence of clustering (Figure~\ref{fig:fig2}(H)--(K))~\citep{Agnew2014,Simpson2013}.

As we demonstrate in Section~\ref{sec:results}, our methods generate accurate ABC posteriors for inference on the discrete problem for a range of biologically relevant parameter regimes, including those where the continuum-limit approximation is poor. In this respect we demonstrate a novel use of approximations that qualitatively respond to changes in parameters in a similar way to the full exact stochastic model. 

%describe Lattice based random walk and 

\section{Methods}
\label{sec:methods}
In this section, we present details of two new algorithms for the acceleration of ABC inference for expensive stochastic models when an appropriate approximation is available. First, we present essential background in ABC inference and sequential Monte Carlo (SMC) samplers for ABC~\citep{Sisson2007,Toni2009}. We then describe our extensions to SMC samplers for ABC and provide numerical examples of our approaches using topical examples from ecology and cell biology.
%While the ABC rejection sampler (Algorithm~\ref{alg:abc_rej}) is the most fundamental ABC sampler, very low acceptance probabilities along with expensive stochastic simulations renders ABC rejection sampling impractical~\citep{Sisson2018,Warne2019b}. Due to the efficacy of ABC ideas, many new techniques have been proposed to improve ABC sampler efficiency, including: Markov chain Monte Carlo~\citep{Marjoram2003}; sequential Monte Carlo ~\citep{Sisson2007,Toni2009}; lazy ABC~\citep{Prangle2016}; approximate ABC~\citep{Buzbas2015}; multilevel ABC~\citep{Jasra2019,Lester2018,Warne2018} and multifidelity ABC~\citep{Prescott2018}. In this work we focus on sequential Monte Carlo samplers for ABC (SMC-ABC)~\citep{Sisson2007,Toni2009}.
%We provide essential background in SMC-ABC as this forms the basis of our methods. We then derive two new algorithms for ABC inference. 

\subsection{Sequential Monte Carlo for Approximate Bayesian computation}

Bayesian analysis techniques are powerful tools for the quantification of uncertainty in parameters, models and predictions~\citep{Gelman2014}. 
%Given observations, $\dat\in \mathbb{D}$, and a proposed model parameterized by the vector $\paramvec \in \paramspace$, then Bayes' Theorem states
%\begin{equation*}
%\CondPDF{\paramvec}{\dat} = \frac{\like{\paramvec}{\dat}\PDF{\paramvec}}{\int_{\paramspace}\like{\paramvec}{\dat}\PDF{\paramvec}\, \text{d}\paramvec},
%\label{eq:bayes}
%\end{equation*}  
%where the \emph{prior}, $\PDF{\paramvec}$, and the \emph{likelihood}, $\like{\paramvec}{\dat}$, represent assumptions on the parameters and model, respectively. The \emph{posterior}, $\CondPDF{\paramvec}{\dat}$, quantifies parameter estimates and uncertainty after conditioning on the observations.
Unfortunately, for many stochastic models of practical interest, the likelihood function is intractable. ABC methods replace likelihood evaluation with an approximation based on stochastic simulations of the proposed model, this is captured directly in \emph{ABC rejection sampling}~\citep{Pritchard1999,Tavare1997} where $\mathcal{M}$ samples are generated from an approximate posterior, denoted by $\CondPDF{\paramvec}{\discrep{\dat}{\simdat} \leq \epsilon}$. Here $\dat_s \sim \simProc{\dat}{\paramvec}$ is a data generation process based on simulation of the model, $\discrep{\dat}{\dat_s}$ is a discrepancy metric, and $\epsilon$ is the discrepancy threshold. The resulting accepted parameter samples are distributed according to $\CondPDF{\paramvec}{\discrep{\dat}{\simdat} \leq \epsilon} \to \CondPDF{\paramvec}{\dat}$ as $\epsilon \to 0$.
%\begin{algorithm}
%	\caption{ABC rejection sampling}
%	\begin{algorithmic}[1]
%		\For{ $i=1,\ldots, \mathcal{M}$}
%		\Repeat
%		\State Sample prior, $\paramvec^* \sim \PDF{\paramvec}$.
%		\State Generate data, $\simdat \sim \simProc{\dat}{\paramvec^*}$.
%		\Until{$\discrep{\dat}{\simdat} \leq \epsilon$}
%		\State Set $\paramvec^{i} \leftarrow \paramvec^{*}$.
%		\EndFor
%	\end{algorithmic}
%	\label{alg:abc_rej}
%\end{algorithm}

The average acceptance probability of a proposed parameter sample $\paramvec^*$ is $\mathcal{O}(\epsilon^{d})$~\citep{Fearnhead2012}, where $d$ is the dimensionality of the data space, $\mathbb{D}$. This renders rejection sampling computationally expensive or even completely prohibitive, especially for high-dimensional parameter spaces~\citep{Marjoram2003,Sisson2007}. Summary statistics can reduce the data dimensionality, however, they will often incur information loss~\citep{Barnes2012,Blum2013,Fearnhead2012}. However, strategies including regression adjustment and marginal adjustment strategies can improve the accuracy of dimension reductions~\citep{Beaumont2002,Nott2014}. 

%\subsection{Sequential Monte Carlo for ABC}

In the SMC-ABC method, importance resampling is applied to a sequence of $R$ ABC posteriors with discrepancy thresholds $ \epsilon_1 > \cdots > \epsilon_R$, with $\epsilon_R$ indicating the target ABC posterior. 
Given $\mathcal{M}$ weighted samples $\{(\paramvec^i,w^i)\}_{i=1}^\mathcal{M}$, called particles, from the prior $\PDF{\paramvec}$, particles are filtered through each ABC posterior using three main steps for each particle $\paramvec^i$: (i) the particle is perturbed using a proposal kernel density $\Kernelsub{\paramvec}{\paramvec^i}{r}$; (ii) an accept/reject step is performed; and (iii) importance weights are updated.
Once all particles have been updated and reweighted, resampling of particles is performed to avoid particle degeneracy. For reference, the SMC-ABC algorithm as initially developed by \citet{Sisson2007} and \citet{Toni2009} is given in Algorithm~\ref{alg:abc_smc}. The number of particles, $\mathcal{M}$, and the number of intermediate distributions, $R$, influence the accuracy and performance, respectively, of the sampler. Setting $\mathcal{M}$ too small can lead to large estimator variability and particle degeneracy, and setting $R$ too small leads to large divergence between successive distributions that can result in high rejection rates.
\begin{algorithm}[h]
	\caption{SMC-ABC}
	\begin{algorithmic}[1]
		\State Initialize $\paramvec_0^{i} \sim \PDF{\paramvec}$ and $w_0^{i} = 1 / \mathcal{M}$, for $i = 1,\ldots, \mathcal{M}$;
		\For{$r = 1,\ldots, R$}
		\For{$i = 1,\ldots, \mathcal{M}$}
		\Repeat
		\State Draw $\paramvec^*$ from $\{\paramvec_{r-1}^j\}_{j=1}^\mathcal{M}$ according to the probability mass function
		\begin{displaymath}
		\Prob{\paramvec^* = \paramvec_{r-1}^j} = \frac{w_{r-1}^j}{\sum_{k=1}^{\mathcal{M}} w_{r-1}^{k}}, \quad \text{for}\, j = 1,\ldots, \mathcal{M}; 
		\end{displaymath}
		\State Sample transition kernel, $\paramvec^{**} \sim \Kernelsub{\paramvec}{\paramvec^*}{r}$;
		\State Generate data, $\simdat \sim \simProc{\dat}{\paramvec^{**}}$;
		\Until{$\discrep{\dat}{\simdat} \leq \epsilon_r$}
		\State Set $\paramvec_{r}^{i} \leftarrow \paramvec^{**}$;
		\State Set $w_r^{i} \leftarrow \PDF{\paramvec_r^{i}}/\left[{\sum_{j=1}^{\mathcal{M}} w_{r-1}^{j} \Kernelsub{\paramvec_r^{i}}{\paramvec_{r-1}^{j}}{r}}\right]$;
		\EndFor
		\State Resample weighted particles, $\{(\paramvec_{r}^{i},w_r^{i})\}_{i=1}^{\mathcal{M}}$, with replacement;
		\State Set $w_r^{i} \leftarrow 1/{\mathcal{M}}$ for all $i = 1,\ldots, \mathcal{M}$;
		\EndFor
	\end{algorithmic}
	\label{alg:abc_smc}
\end{algorithm}
Note that throughout all algorithms used in this manuscript, we assume that the initial set of weighted particles, $\{\paramvec_0^i,w_0^i\}_{i=1}^\mathcal{M}$, are independent, identically distributed samples from the prior, $\PDF{\paramvec}$, and therefore have uniform weight, $w_0^i = 1/\mathcal{M}$, for all $i = 1,\ldots, \mathcal{M}$. However, the methods are general enough to deal with prior distributions that require importance sampling to draw truly weighted particles.

For a fixed choice of $R$, efficient use of SMC-ABC depends critically on the selection of appropriate proposal kernels and threshold sequences. The process of sampling at the target threshold, $\epsilon_r$, given the weights of the previous threshold, $\epsilon_{r-1}$, is described by~\cite{Filippi2013}
\begin{equation}
\xi_{r}\left(\paramvec_r \mid \dat \right) = \frac{1}{a_{r,r-1}}\int_{\paramspace}\int_{\mathbb{D}} \ind{B(\dat,\epsilon_{r})}{\simdat}\simProc{\simdat}{\paramvec_{r}}\Kernelsub{\paramvec_{r}}{\paramvec_{r-1}}{r}w(\paramvec_{r-1})\,\text{d}\simdat \text{d}\paramvec_{r-1}, 
\label{eq:proc_r}
\end{equation}
where the data space, $\mathbb{D}$, has dimensionality $d$, $B(\dat,\epsilon_r)$ is a $d$-dimensional ball centered on the data with radius $\epsilon_r$, and $\ind{A}{x}$ denotes the indicator function with $\ind{A}{x} = 1$ if $x \in A$, otherwise $\ind{A}{x} = 0$. The normalization constant, $a_{r,r-1}$, can be interpreted as the average acceptance probability across all particles. We see this by noting that \eqref{eq:proc_r} can be reduced to 
\begin{equation}
\xi_{r}\left(\paramvec_r \mid \dat \right) = \frac{\eta_{r-1}(\paramvec_r) \CondProb{\simdat \in B(\dat,\epsilon_{r})}{\paramvec_{r}}}{a_{r,r-1}}.
\label{eq:accep}
\end{equation}
Here the distribution $\eta_{r-1}(\paramvec_r)$ represent the proposal mechanism,
\begin{equation}
\eta_{{r-1}}(\paramvec_{r}) = \sum_{j=1}^\mathcal{M}\Kernelsub{\paramvec_r}{\paramvec_{r-1}^j}{r}w(\paramvec_{r-1}),
\label{eq:exactprop}
\end{equation} and $\CondProb{\simdat \in B(\dat,\epsilon_{r})}{\paramvec_{r}}$ is the probability that simulated data is within $\epsilon_r$ of the data $\dat$ given a parameter value $\paramvec_r$. Therefore, the normalizing constant is
\begin{equation}
a_{r,r-1} = \E{\CondProb{\simdat \in B(\dat,\epsilon_{r})}{\paramvec_{r}}},
\label{eq:averageAccept}
\end{equation}
that is, $a_{r,r-1}$ is the average acceptance probability. 

From a computational perspective, the goal is to choose $\eta_{r-1}(\paramvec_r)$ to maximize $a_{r,r-1}$. However, this would not necessarily result in a $\xi_{r}\left(\paramvec_r \mid \dat \right)$ that is an accurate approximation to the true target ABC posterior $\CondPDF{\paramvec_r}{\discrep{\dat}{\simdat} \leq \epsilon_r}$. To achieve this goal, we require $\eta_{r-1}(\paramvec_r)$ such that the Kullback-Leibler divergence~\citep{Kullback1951}, $D_{\text{KL}}\left(\xi_{r}\left(\cdot \mid \dat \right);\CondPDF{\cdot}{\discrep{\dat}{\simdat} \leq \epsilon_r} \right)$, is minimized. 
\citet{Beaumont2009} and \citet{Filippi2013} demonstrate how the latter goal provides insight into how to optimally choose $\eta_{r-1}(\paramvec_r)$. The key is to note that $D_{\text{KL}}\left(\xi_{r}\left(\cdot \mid \dat \right);\CondPDF{\cdot}{\discrep{\dat}{\simdat} \leq \epsilon_r} \right)$ can be decomposed as follows,
\begin{align}
\begin{split}
D_{\text{KL}} \left(\xi_{r}\left(\cdot \mid \dat \right);\CondPDF{\cdot}{\discrep{\dat}{\simdat} \leq \epsilon_r} \right) 
%D_{\text{KL}} & \left(\xi_{r}\left(\cdot \mid \dat \right);\CondPDF{\cdot}{\discrep{\dat}{\simdat} \leq \epsilon_r} \right) = \int_{\paramspace} \CondPDF{\paramvec_r}{\discrep{\dat}{\simdat} \leq \epsilon_r} \log_e\left[\frac{\CondPDF{\paramvec_r}{\discrep{\dat}{\simdat} \leq \epsilon_r}}{\xi_{{r}}(\paramvec_r\mid \dat)}\right]\, \text{d}\paramvec_r \notag \\
&= D_{\text{KL}}\left(\eta_{r-1}\left(\cdot\right);\CondPDF{\cdot}{\discrep{\dat}{\simdat} \leq \epsilon_r} \right) \\ &\quad- E(\paramvec_r) + \log_e a_{r,r-1}, 
\end{split}
\label{eq:decomp}
\end{align} 
where
$E(\paramvec_r) = \E{\log_e\left(\int_{\mathbb{D}} \ind{B(\dat,\epsilon_{r})}{\simdat}\simProc{\simdat}{\paramvec_{r}}\,\text{d}\simdat\right)}$
is independent of $\eta_{r-1}(\paramvec_{r})$. By rearranging \eqref{eq:decomp}, we obtain
\begin{equation*}
D_{\text{KL}}\left(\eta_{r-1}\left(\cdot\right);\CondPDF{\cdot}{\discrep{\dat}{\simdat} \leq \epsilon_r}\right) = D_{\text{KL}}\left(\xi_{r}\left(\cdot \mid \dat \right);\CondPDF{\cdot}{\discrep{\dat}{\simdat} \leq \epsilon_r} \right) + E(\paramvec_r) - \log_e a_{r,r-1}.
\end{equation*}
That is, minimizing $D_{\text{KL}}\left(\eta_{r-1}\left(\cdot\right);\CondPDF{\cdot}{\discrep{\dat}{\simdat} \leq \epsilon_r}\right)$ is equivalent to minimizing\\ $D_{\text{KL}}\left(\xi_{r}\left(\cdot \mid \dat \right);\CondPDF{\cdot}{\discrep{\dat}{\simdat} \leq \epsilon_r} \right)$ and maximizing $a_{r,r-1}$ simultaneously.
Therefore, any proposal mechanism that is closer, in the Kullback-Liebler sense, to  $\CondPDF{\cdot}{\discrep{\dat}{\simdat} \leq \epsilon_r}$ is more efficient.

We apply the optimal adaptive scheme of~\cite{Beaumont2009} and~\cite{Filippi2013} for multivariate Gaussian proposals. That is, we set
\begin{equation*}
\Kernelsub{\paramvec_r}{\paramvec_{r-1}}{r} = \frac{1}{\sqrt{(2\pi)^n\det(\Sigma)}}\exp\left(-(\paramvec_r-\paramvec_{r-1})^\text{T}\Sigma^{-1}(\paramvec_r-\paramvec_{r-1})/2\right),
\end{equation*}  
where $n$ is the dimensionality of parameter space, $\paramspace$, and 
\begin{equation*}
\Sigma = \frac{2}{\mathcal{M}-1} \sum_{i=1}^{\mathcal{M}} (\paramvec_{r-1}^i - \mu_{r-1}) (\paramvec_{r-1}^i - \mu_{r-1})^{\text{T}} \quad\text{with} \quad \mu_{r-1} = \frac{1}{\mathcal{M}} \sum_{i=1}^{\mathcal{M}} \paramvec_{r-1}^i.
\end{equation*}

%In an adaptive SMC scheme, the proposal kernel, $\Kernelsub{\paramvec^*}{\paramvec}{r}$, is chosen such that the efficiency is optimal under certain assumptions on the kernel and target distribution families (Supplementary Material). In this work, we apply such an adaptive scheme~\citep{Beaumont2009,Drovandi2011,Filippi2013} that seeks to select an optimal proposal kernel, $\Kernelsub{\paramvec_r}{\paramvec_{r-1}}{r}$ based on the weighted samples from the previous iteration $\{(\paramvec_{r-1}^{i},w_{r-1}^{i})\}_{i=1}^{\mathcal{M}}$  (Supplementary Material).
%\todo{we base our work on an extension of SMC for ABC~\citep{Sisson2007,Toni2009} and leave adaptive methods for future work~}
\subsection{Preconditioning SMC-ABC}

Consider a fixed sequence of ABC posteriors for the stochastic model inference problem, $\{\CondPDF{\paramvec}{\discrep{\dat}{\simdat} \leq \epsilon_r}\}_{r=1}^R$. We want to apply SMC-ABC (Supplementary Material) to efficiently sample from this sequence with adaptive proposal kernels, $\{\Kernelsub{\paramvec^*}{\paramvec}{r}\}_{r=1}^R$~\citep{Beaumont2009,Filippi2013}. Our method exploits an approximate model to further improve the average acceptance probability. 

\subsubsection{Algorithm development}

Say we have a set of weighted particles that represent the ABC posterior at threshold $\epsilon_{r-1}$ using the stochastic model, that is, $\{(\paramvec_{r-1}^i,w_{r-1}^i)\}_{i=1}^\mathcal{M} \approx \CondPDF{\paramvec_{r-1}}{\discrep{\dat}{\simdat} \leq \epsilon_{r-1}}$. Now, consider applying the next importance resampling step using an approximate data generation step, $\approxsimdat \sim \approxsimProc{\approxsimdat}{\paramvec}$, where $\approxsimProc{\approxsimdat}{\paramvec}$ is the simulation process of an approximate model\footnote{Throughout, the overbar tilde notation, e.g. $\tilde{x}$, is used to refer to the ABC entities related to the approximate model, whereas quantities without the overbar tilde notation, e.g. $x$, are used to refer fo the ABC entities related to the exact model.}. Furthermore, assume the computational cost of simulating the approximate model, $\text{Cost}(\approxsimdat)$, is significantly less than the computational cost of the exact model, $\text{Cost}(\simdat)$, that is, $\text{Cost}(\approxsimdat)/\text{Cost}(\simdat) \ll 1$. The result will be a new set of particles that represent the ABC posterior at threshold $\epsilon_{r}$ using this approximate model, denoted $\{(\approxparamvec_{r}^i,\tilde{w}_{r}^i)\}_{i=1}^\mathcal{M} \approx \approxCondPDF{\approxparamvec_{r}}{\discrep{\dat}{\approxsimdat} \leq\epsilon_{r}}$. As noted in the examples in Section~\ref{sec:intro}, approximate models are not always valid. This implies that $ \approxCondPDF{\approxparamvec_{r}}{\discrep{\dat}{\approxsimdat} \leq\epsilon_{r}}$ is always biased and will not in general converge to $\CondPDF{\paramvec}{\dat}$ as $\epsilon_{r} \to 0$. However, since $\text{Cost}(\approxsimdat)/\text{Cost}(\simdat) \ll 1$, it is computationally inexpensive to compute the distribution  
\begin{equation}
\tilde{\eta}_{{r}}(\paramvec_{r}) = \sum_{j=1}^\mathcal{{M}}\approxKernelsub{\paramvec_r}{\approxparamvec_{r}^j}{r}\tilde{w}_{r}^j,
\label{eq:approxprop}
\end{equation}
in comparison to computing $\eta_{{r-1}}(\paramvec_{r})$ (\eqref{eq:exactprop}). In \eqref{eq:approxprop}, the proposal kernel $\approxKernelsub{\paramvec_r}{\approxparamvec_{r}^j}{r}$ is possibly distinct from the $\Kernelsub{\paramvec_r}{\paramvec_{r-1}^j}{r}$ used in $\eta_{{r-1}}(\paramvec_{r})$ (\eqref{eq:exactprop}). To improve the efficiency of the sampling process we simply require
\begin{equation}
D_{\text{KL}}(\eta_{{r-1}}(\cdot) ;\CondPDF{\cdot}{\discrep{\dat}{\simdat} \leq\epsilon_r}) > D_{\text{KL}}(\tilde{\eta}_{{r}}(\cdot) ;\CondPDF{\cdot}{\discrep{\dat}{\approxsimdat}\leq\epsilon_r}), \label{eq:effcond}
\end{equation}
for $\tilde{\eta}_{{r}}(\paramvec_{r})$ (\eqref{eq:approxprop}) to be more efficient as a proposal mechanism compared with $\eta_{{r-1}}(\paramvec_{r})$ (\eqref{eq:exactprop}). Provided the condition $\text{Cost}(\approxsimdat)/\text{Cost}(\simdat)~\ll~1$ holds, any improvements in sampling efficiency will translate directly into computational performance improvements. That is, it does not matter that $\approxCondPDF{\approxparamvec_{r}}{\discrep{\dat}{\approxsimdat} \leq\epsilon_{r}}$ is biased, it just needs to be less biased than $\CondPDF{\paramvec_{r-1}}{\discrep{\dat}{\simdat} \leq\epsilon_{r-1}}$ and computationally inexpensive.   

This idea yields an intuitive new algorithm for SMC-ABC; proceed through the sequential sampling of $\{\CondPDF{\paramvec}{\discrep{\dat}{\simdat}\leq\epsilon_r}\}_{r=1}^R$ by applying two resampling steps for each iteration. The first moves the particles from acceptance threshold $\epsilon_{r-1}$ to $\epsilon_{r}$ using the computationally inexpensive approximate model, and the second corrects for the bias between $\approxCondPDF{\approxparamvec_{r}}{\discrep{\dat}{\approxsimdat}\leq\epsilon_{r}}$ and $\CondPDF{\paramvec_{r}}{\discrep{\dat}{\simdat}\leq\epsilon_{r}}$ using the expensive stochastic model, but at an improved acceptance rate. 
Since the intermediate distribution acts on the proposal mechanism to accelerate the convergence time of SMC-ABC, we denote the sequence $\{\approxCondPDF{\approxparamvec_{r}}{\discrep{\dat}{\approxsimdat}\leq\epsilon_{r}}\}_{r=1}^R$ as the preconditioner distribution sequence. The algorithm, called \emph{preconditioned SMC-ABC} (PC-SMC-ABC), is given in Algorithm~\ref{alg:abc_pcsmc}.  We note that similar notions of preconditioning with approximation informed proposals have been applied in the context of Markov chain Monte Carlo samplers~\citep{Parno2018}. However, to the best of our knowledge, our approach represents the first application of preconditioning ideas to SMC-ABC. 
\newpage
\begin{algorithm}[h]
	\caption{Preconditioned SMC-ABC}
	\begin{algorithmic}[1]
		\State Initialize $\paramvec_0^{i} \sim \PDF{\paramvec}$ and $w_0^{i} = 1/{\mathcal{M}}$, for $i = 1,\ldots, \mathcal{M}$;
		\For{$r = 1,\ldots, R$}
		\For{$i = 1,\ldots, \mathcal{M}$}
		\Repeat
		\State Draw $\paramvec^*$ from $\{\paramvec_{r-1}^j\}_{j=1}^\mathcal{M}$ according to the probability mass function
		\begin{displaymath}
		\Prob{\paramvec^* = \paramvec_{r-1}^j} = \frac{w_{r-1}^j}{\sum_{k=1}^{\mathcal{M}} w_{r-1}^{k}}, \quad \text{for}\, j = 1,\ldots, \mathcal{M}; 
		\end{displaymath}
		%\State Set $\paramvec^* \leftarrow \paramvec_{r-1}^j$ with probability $w_{r-1}^j/\left[{\sum_{k=1}^{\mathcal{M}} w_{r-1}^{k}}\right]$;
		\State Sample transition kernel, $\approxparamvec^{**} \sim \Kernelsub{\approxparamvec}{\paramvec^*}{r}$;
		\State Generate data, $\approxsimdat \sim \approxsimProc{\dat}{\approxparamvec^{**}}$;
		\Until{$\discrep{\dat}{\approxsimdat} \leq \epsilon_r$}
		\State Set $\approxparamvec_{r}^{i} \leftarrow \approxparamvec^{**}$;
		\State Set $\tilde{w}_r^{i} \leftarrow \PDF{\approxparamvec_r^{i}}/ \left[{ \sum_{j=1}^{\mathcal{M}} w_{r-1}^{j} \Kernelsub{\approxparamvec_r^{i}}{\paramvec_{r-1}^{j}}{r}}\right]$;
		\EndFor
		\For{$i = 1,\ldots, \mathcal{M}$}
		\Repeat
		\State Draw $\approxparamvec^*$ from $\{\approxparamvec_{r-1}^j\}_{j=1}^\mathcal{M}$ according to the probability mass function
		\begin{displaymath}
		\Prob{\approxparamvec^* = \approxparamvec_{r-1}^j} = \frac{\tilde{w}_{r-1}^j}{\sum_{k=1}^{\mathcal{M}} \tilde{w}_{r-1}^{k}}, \quad \text{for}\, j = 1,\ldots, \mathcal{M}; 
		\end{displaymath}
		%\State Set $\approxparamvec^* \leftarrow \approxparamvec_{r}^j$ with probability $\tilde{w}_{r}^j/\left[{\sum_{k=1}^{\mathcal{M}} \tilde{w}_{r}^{k}}\right]$;
		\State Sample transition kernel, $\paramvec^{**} \sim \approxKernelsub{\paramvec}{\approxparamvec^*}{r}$;
		\State Generate data, $\simdat \sim \simProc{\dat}{\paramvec^{**}}$;
		\Until{$\discrep{\dat}{\simdat} \leq \epsilon_r$}
		\State Set $\paramvec_{r}^{i} \leftarrow \paramvec^{**}$;
		\State Set $w_r^{i} \leftarrow \PDF{\paramvec_r^{i}}/ \left[{\sum_{j=1}^{\mathcal{M}} \tilde{w}_{r}^{j} \approxKernelsub{\paramvec_r^{i}}{\approxparamvec_{r}^{j}}{r}}\right]$;
		\EndFor
		\State Resample weighted particles, $\{(\paramvec_{r}^{i},w_r^{i})\}_{i=1}^{\mathcal{M}}$, with replacement;
		\State Set $w_r^{i} \leftarrow 1/{\mathcal{M}}$ for all $i = 1,\ldots, \mathcal{M}$;
		\EndFor
	\end{algorithmic}
	\label{alg:abc_pcsmc}
\end{algorithm}
One particular advantage of the PC-SMC-ABC method, that is demonstrated in the next section, is that it is unbiased. Effectively, one can consider PC-SMC-ABC as standard SMC-ABC method with a specialized proposal mechanism based on the preconditioner distribution. This means that PC-SMC-ABC is completely general, as discussed in Section~\ref{sec:discuss}, and is independent of the specific stochastic models that we consider here.
%\FloatBarrier
This property of unbiasedness holds even for cases where the approximate model is a poor approximation of the forward dynamics of the model. However, the closer that $\tilde{\eta}_r(\paramvec_{r})$ is to $\CondPDF{\paramvec_r}{\discrep{\dat}{\simdat} \leq \epsilon_r}$ the better the performance improvement will be, as we demonstrate in Section~\ref{sec:results}.

%We have yet to find an example, even in extreme cases where the approximate model is a very poor description of the stochastic model, that result in decreased performance (Section~\ref{sec:results}).
%However, it is nontrivial to test \emph{a priori} that the essential condition given in \eqref{eq:effcond} holds in practice. Certainly, in cases where the approximate model provides an accurate description of the stochastic model, this approach works very well. Conversely, one could also consider cases with extreme bias from the approximate model; in such scenarios it is likely that there would exist a threshold, $\epsilon^*$, such that \eqref{eq:effcond} is violated. In such cases, our PC-SMC-ABC method is not guaranteed to perform better than SMC-ABC on the discrete model. 

\FloatBarrier

\subsection{Moment-matching SMC-ABC}
\label{sec:mmsmcabc}
%TODO remove mathcalM bar and tilde...
The PC-SMC-ABC method is a promising modification to SMC-ABC that can accelerate inference for expensive stochastic models without introducing bias. However, other approaches can be used  to obtain further computational improvements. Here, we consider an alternate approach to utilizing approximate models that aims to get the most out of a small sample of expensive stochastic simulations. Unlike PC-SMC-ABC, this method is generally biased, but it has the advantage of yielding a small and fixed computational budget.
Specifically, we define a parameter $\alpha \in [0,1]$, such that $1/\alpha$ is the target computational speedup, for example, $\alpha = 1/10$ should result in approximate $10$ times speedup. We apply the SMC-ABC method using $\tilde{\mathcal{M}} = \floor{(1-\alpha)\mathcal{M}}$ particles based on the approximate model, and then use $\hat{\mathcal{M}} = \ceil{\alpha \mathcal{M}}$ particles based on the stochastic model to construct a hybrid population of  $\mathcal{M} = \hat{\mathcal{M}} + \tilde{\mathcal{M}}$ particles that will represent the final inference on the stochastic model. The key idea is that we use the $\hat{\mathcal{M}}$ particles of the expensive stochastic model to inform a transformation on the $\tilde{\mathcal{M}}$ particles of the approximation such that they the emulate particles of expensive stochastic model. Here, $\floor{\cdot}$ and $\ceil{\cdot}$ are, respectively,  the floor and ceiling functions.

\subsubsection{Algorithm development}
Assume that we have applied SMC-ABC to sequentially sample $\tilde{\mathcal{M}}$ particles through the ABC posteriors from the approximate model, $\{\approxCondPDF{\approxparamvec_{r}}{\discrep{\dat}{\approxsimdat}\leq\epsilon_r}\}_{r=1}^R$, with $\epsilon_R = \epsilon$. For the sake of the derivation, say that for all $r \in [1,R]$ we have available the mean vector, $\boldsymbol{\mu}_r$, and the covariance matrix, $\boldsymbol{\Sigma}_r$, of the ABC posterior $\CondPDF{\paramvec_{r}}{\discrep{\dat}{\simdat}\leq\epsilon_r}$ under the stochastic model. In this case, we use particles $\approxparamvec_{r}^1,\ldots,\approxparamvec_{r}^{\tilde{\mathcal{M}}}\sim \approxCondPDF{\approxparamvec_{r}}{\discrep{\dat}{\approxsimdat}\leq\epsilon_r}$ to emulate particles $\paramvec_{r}^1,\ldots,\paramvec_{r}^{\hat{\mathcal{M}}} \sim \CondPDF{\paramvec_{r}}{\discrep{\dat}{\simdat}\leq\epsilon_r}$ by using the moment matching transform~\citep{Lei2011,Sun2016}
\begin{equation}
\paramvec_r^i = \bvec{L}_r\left[\tilde{\bvec{L}}_r^{-1}\left(\approxparamvec_r^i - \tilde{\boldsymbol{\mu}}_r\right) \right] + \boldsymbol{\mu}_r, \quad i = 1,2,\ldots, \tilde{\mathcal{M}}, \label{eq:mmtf}
\end{equation}
where $\tilde{\boldsymbol{\mu}}_r$ and $\tilde{\boldsymbol{\Sigma}}_r$ are the empirical mean vector and convariance matrix of particles $\approxparamvec_{r}^1,\ldots,\approxparamvec_{r}^{\tilde{\mathcal{M}}}$, $\bvec{L}_r$ and $\tilde{\bvec{L}}_r$ are lower triangular matrices obtained through the Cholesky factorization~\citep{Press1997} of $\boldsymbol{\Sigma}_r$ and $\tilde{\boldsymbol{\Sigma}}_r$, respectively, and $\tilde{\bvec{L}}_r^{-1}$ is the matrix inverse of $\tilde{\bvec{L}}_r$. This transform will produce a collection of particles that has a sample mean vector of $\boldsymbol{\mu}_r$ and covariance matrix $\boldsymbol{\Sigma}_r$. That is, the transformed sample matches the ABC posterior under the stochastic model up to the first two moments. In Section~\ref{sec:results}, we demonstrate that matching two moments is sufficient for the problems we investigate here, however, in principle we could extend this matching to higher order moments if required. For discussion on the advantages and disadvantages of matching higher moments, see Section~\ref{sec:discuss}.   

In practice, it would be rare that $\boldsymbol{\mu}_r$ and $\boldsymbol{\Sigma}_r$ are known. If $\CondPDF{\paramvec_{r-1}}{\discrep{\dat}{\simdat}\leq\epsilon_{r-1}}$ is available, then we can use importance resampling to obtain $\hat{\mathcal{M}}$ particles, $\paramvec_{r}^1,\ldots,\paramvec_{r}^{\hat{\mathcal{M}}}$, from $\CondPDF{\paramvec_{r}}{\discrep{\dat}{\simdat}\leq\epsilon_{r}}$, that is, we perform a step from SMC-ABC using the expensive stochastic model. We can then use the unbiased estimators
\begin{equation}
\hat{\boldsymbol{\mu}}_r = \frac{1}{\hat{\mathcal{M}}} \sum_{i=1}^{\hat{\mathcal{M}}} \paramvec_{r}^i \quad\text{and} \quad \hat{\boldsymbol{\Sigma}}_r = \frac{1}{\hat{\mathcal{M}}-1} \sum_{i=1}^{\hat{\mathcal{M}}} (\paramvec_{r}^i - \hat{\boldsymbol{\mu}}_r) (\paramvec_{r}^i - \hat{\boldsymbol{\mu}}_r)^{\text{T}}, \label{eq:estmusig}
\end{equation}
to obtain estimates of $\boldsymbol{\mu}_r$ and $\boldsymbol{\Sigma}_r$. Substituting \eqref{eq:estmusig} into \eqref{eq:mmtf} gives an approximate transform 
\begin{equation}
\hat{\paramvec}_r^i = \hat{\bvec{L}}_r\left[\tilde{\bvec{L}}_r^{-1}\left(\approxparamvec_r^i - \tilde{\boldsymbol{\mu}}_r\right) \right] + \hat{\boldsymbol{\mu}}_r, \quad i = 1,2,\ldots, \tilde{\mathcal{M}}, \label{eq:mmtfapprox}
\end{equation}
where $\hat{\boldsymbol{\Sigma}}_r = \hat{\bvec{L}}_r\hat{\bvec{L}}_r^{\text{T}}$.
This enables us to construct an estimate of $\CondPDF{\paramvec_{r}}{\discrep{\dat}{\simdat}\leq\epsilon_r}$ by applying the moment-matching transform (\eqref{eq:mmtfapprox})  to the particles $\approxparamvec_{r}^1,\ldots,\approxparamvec_{r}^{\tilde{\mathcal{M}}}$ then pooling the transformed particles $\hat{\paramvec}_{r}^1,\ldots,\hat{\paramvec}_{r}^{\tilde{\mathcal{M}}}$ with the particles $\paramvec_{r}^1,\ldots,\paramvec_{r}^{\hat{\mathcal{M}}}$ that were used in the estimates $\hat{\boldsymbol{\mu}}_r$ and $\hat{\boldsymbol{\Sigma}}_r$. The goal of the approximate transform application is for the transforms particles $\hat{\paramvec}_{r}^1,\ldots,\hat{\paramvec}_{r}^{\tilde{\mathcal{M}}}$ to be more accurate in higher moments due, despite only matching the first two moments (see Section~\ref{sec:selalpha} for numerical justification of this property for some specific examples). This results in an approximation of $\CondPDF{\paramvec_{r}}{\discrep{\dat}{\simdat}\leq\epsilon_r}$ using a set of $\mathcal{M}$ particles $\paramvec_{r}^1,\ldots,\paramvec_{r}^{\mathcal{M}}$ with $\paramvec_{r}^{i+\hat{\mathcal{M}}} = \hat{\paramvec}_{r}^i$ where $1 \leq i \leq \tilde{\mathcal{M}}$.

\begin{algorithm}[h]
	\caption{Moment-matching SMC-ABC}
	\begin{algorithmic}[1]
		\State Given $\alpha \in [0,1]$, initialize $\hat{\mathcal{M}}=\ceil{\alpha \mathcal{M}}$ and $\tilde{\mathcal{M}}= \floor{(1-\alpha)\mathcal{M}}$;
		\State Initialize $\approxparamvec_0^{i} \sim \PDF{\paramvec}$ and $\tilde{w}_0^{i} = 1/{\tilde{\mathcal{M}}}$, for $i = 1,\ldots, \tilde{\mathcal{M}}$; 
		\State Initialize $\paramvec_0^{i} \sim \PDF{\paramvec}$ and $w_0^{i} = 1/{\mathcal{M}}$, for $i = 1,\ldots, \mathcal{M}$;
		\State Apply SMC-ABC to generate the sequence of approximate particles  $\{(\approxparamvec_{1}^{i},\tilde{w}_1^{i})\}_{i=1}^{\tilde{\mathcal{M}}}$, $\{(\approxparamvec_{2}^{i},\tilde{w}_2^{i})\}_{i=1}^{\tilde{\mathcal{M}}},\ldots, \{(\approxparamvec_{R}^{i},\tilde{w}_R^{i})\}_{i=1}^{\tilde{\mathcal{M}}}$
		\For{$r = 1,\ldots, R$}
		%\State Resample weighted particles, $\{(\approxparamvec_{t}^{i},\tilde{w}_t^{i})\}_{i=1}^{\floor{(1-\alpha)M}}$, with replacement;
		%\State Set $\tilde{w}_t^{i} \leftarrow 1/\floor{(1-\alpha)M}$ for all $i = 1,\ldots, \floor{(1-\alpha)M}$;
		\For{$i = 1,\ldots, \hat{\mathcal{M}}$}
		\Repeat
		\State Draw $\paramvec^*$ from $\{\paramvec_{r-1}^j\}_{j=1}^\mathcal{M}$ according to the  probability mass function
		\begin{displaymath}
		\Prob{\paramvec^* = \paramvec_{r-1}^j} = \frac{w_{r-1}^j}{\sum_{k=1}^{\mathcal{M}} w_{r-1}^{k}}, \quad \text{for} j = 1,\ldots, \mathcal{M}; 
		\end{displaymath}
		%\State Set $\paramvec^* \leftarrow \paramvec_{r-1}^j$ with probability $w_{r}^j/\left[{\sum_{k=1}^{\mathcal{M}} w_{r-1}^{k}}\right]$;
		\State Sample transition kernel, $\paramvec^{**} \sim \Kernelsub{\paramvec}{\paramvec^*}{r}$;
		\State Generate data, $\simdat \sim \simProc{\dat}{\paramvec^{**}}$;
		\Until{$\discrep{\dat}{\simdat} \leq \epsilon_r$}
		\State Set $\paramvec_{r}^{i} \leftarrow \paramvec^{**}$;
		\State Set $w_r^{i} \leftarrow \PDF{\paramvec_r^{i}}/\left[{ \sum_{j=1}^{\mathcal{M}} w_{r-1}^{j} \Kernelsub{\paramvec_r^{i}}{\paramvec_{r-1}^{j}}{r}}\right]$;
		\EndFor
		%\State Resample weighted particles, $\{(\hat{\paramvec}_{t}^{i},\hat{w}_t^{i})\}_{i=1}^{\ceil{\alpha M}}$, with replacement;
		%\State Set $\hat{w}_t^{i} \leftarrow 1/\ceil{\alpha M}$ for all $i = 1,\ldots, \ceil{\alpha M}$;
		\State Estimate means and covariances $\tilde{\boldsymbol{\mu}}_r$, $\tilde{\boldsymbol{\Sigma}}_r$, $\hat{\boldsymbol{\mu}}_r$, and $\hat{\boldsymbol{\Sigma}}_r$;
		\State Compute Cholesky decompositions $\tilde{\boldsymbol{\Sigma}}_r = \tilde{\bvec{L}}_r\tilde{\bvec{L}}_r^\text{T}$ and $\hat{\boldsymbol{\Sigma}}_r = \hat{\bvec{L}}_r\hat{\bvec{L}}_r^\text{T}$
		\For{$i=1,\ldots, \tilde{\mathcal{M}}$}
		\State Set $\paramvec_{r}^{i+\hat{\mathcal{M}}} \leftarrow \hat{\bvec{L}}_r[\tilde{\bvec{L}}_r^{-1}(\approxparamvec_r^{i} - \tilde{\boldsymbol{\mu}}_r)] + \hat{\boldsymbol{\mu}}_r$ and $w_{r}^{i+\hat{\mathcal{M}}} \leftarrow \tilde{w}_r^{i}$;
		\EndFor
		\State Resample weighted particles $\{(\paramvec_{r}^{i},w_r^{i})\}_{r=1}^{\mathcal{M}}$ with replacement;
		\State Set $w_r^{i} \leftarrow 1/\mathcal{M}$ for all $i = 1,\ldots, \mathcal{M}$;
		\EndFor
	\end{algorithmic}
	\label{alg:abc_mmsmc}
\end{algorithm}

This leads to our \emph{moment-matching} SMC-ABC (MM-SMC-ABC) method. First, SMC-ABC inference is applied using the approximate model with $\tilde{\mathcal{M}}$ particles. Then, given $\mathcal{M}$ samples from the prior, $\PDF{\paramvec}$, we can sequentially approximate\\ $\{\CondPDF{\paramvec_r}{\discrep{\dat}{\simdat}\leq\epsilon_r}\}_{r=1}^R$. At each iteration the following steps are performed: (i) generate a small number of particles from $\CondPDF{\paramvec_r}{\discrep{\dat}{\simdat}\leq\epsilon_r}$ using importance resampling and stochastic model simulations; (ii) compute $\hat{\boldsymbol{\mu}}_r$ and $\hat{\boldsymbol{\Sigma}}_r = \hat{\bvec{L}}_r\hat{\bvec{L}}_r^{\text{T}}$; (iii) apply the transform from \eqref{eq:mmtfapprox} to the particles at $\epsilon_{r}$ from the approximate model; (iv) pool the resulting particles with the stochastic model samples; and (v) reweight particles and resample.
The final MM-SMC-ABC algorithm is provided in Algorithm~\ref{alg:abc_mmsmc}. 

The performance of this method depends on the choice of $\alpha$. Note that in Algorithm~\ref{alg:abc_mmsmc}, standard SMC-ABC for the expensive stochastic model is recovered as $\alpha \to 1$\\ (no speedup, inference unbiased), and standard SMC-ABC using the approximate model is recovered as $\alpha \to 0$ (maximum speedup, but inference biased). Therefore we expect there is a choice of $ \alpha \in (0,1)$ that provides an optimal trade-off between computational improvement and accuracy. Clearly, the expected speed improvement is proportional to $1/\alpha$, however, if $\alpha$ is chosen to be too small, then the statistical error in the estimates in \eqref{eq:estmusig} will be too high. We explore this trade-off in detail in Section~\ref{sec:selalpha} and find that $0.05 \leq \alpha \leq 0.2$ seems to give a reasonable result.

\FloatBarrier 
%TODO: rephrase, a bit clunky ~ REB.
%The MM-SMC-ABC method is biased, there is an assumption that the inferences using the approximate model form a good approximation to inferences using the stochastic model once adjusting for the mean vector and convariance matrices. That is, they come from the same family of distributions that are parameterised by the first two moment only. It is unlikely that this holds in general, thus we conclude that MM-SMC-ABC is biased. However, as shown in Section~\ref{sec:results}, this bias is likely to be lower than that which results from using the approximate model alone, and we observe an increase in the computational performance by an order of magnitude.
%TODO: upsell to very general case.

\section{Results}
\label{sec:results}
In this section, we provide numerical examples to demonstrate the accuracy and performance of the PC-SMC-ABC and MM-SMC-ABC methods. First we apply PC-SMC-ABC to a tractable example to demonstrate the mechanisms of the method and provide insight into effective choices of approximate model. The tractable example considered here is inference for an Ornstein--Uhlenbeck process~\citep{Ornstein1930}. 
We then consider two intractable problems based on expensive discrete model. For our first example, we consider the analysis of spatially averaged population growth data. The discrete model used in this instance is relevant in the ecological sciences as it describes  population growth subject to a weak Allee effect~\citep{Taylor2005}. We then analyze data that is typical of \emph{in vitro} cell culture scratch assays in experimental cell biology using a discrete model that leads to the well-studied Fisher-KPP model~\citep{EdelsteinKeshet2005,Murray2002}. In both examples, we present the discrete model and its continuum limit, then compute the full Bayesian posterior for the model parameters using the PC-SMC-ABC (Algorithm~\ref{alg:abc_pcsmc}) and MM-SMC-ABC (Algorithm~\ref{alg:abc_mmsmc}) methods, and compare the results with the SMC-ABC (Supplementary Material) using either the discrete model or continuum limit alone. We also provide numerical experiments to evaluate the effect of the tuning parameter $\alpha$ on the accuracy and performance of the MM-SMC-ABC method.

%TODO: think about speedup as a function of cost of simulations of both models
It is important to clarify that when we refer to the accuracy of our methods, we refer to their ability to sample from the target ABC posterior under the expensive stochastic model. The evaluation of this accuracy requires sampling from the target ABC posterior under the expensive stochastic model using SMC-ABC. As a result, the target acceptance thresholds are chosen to ensure this is computationally feasible. 

\subsection{A tractable example: Ornstein--Uhlenbeck process}
\label{sec:ousde}
\begin{figure}[h!]
	\includegraphics[width=\linewidth]{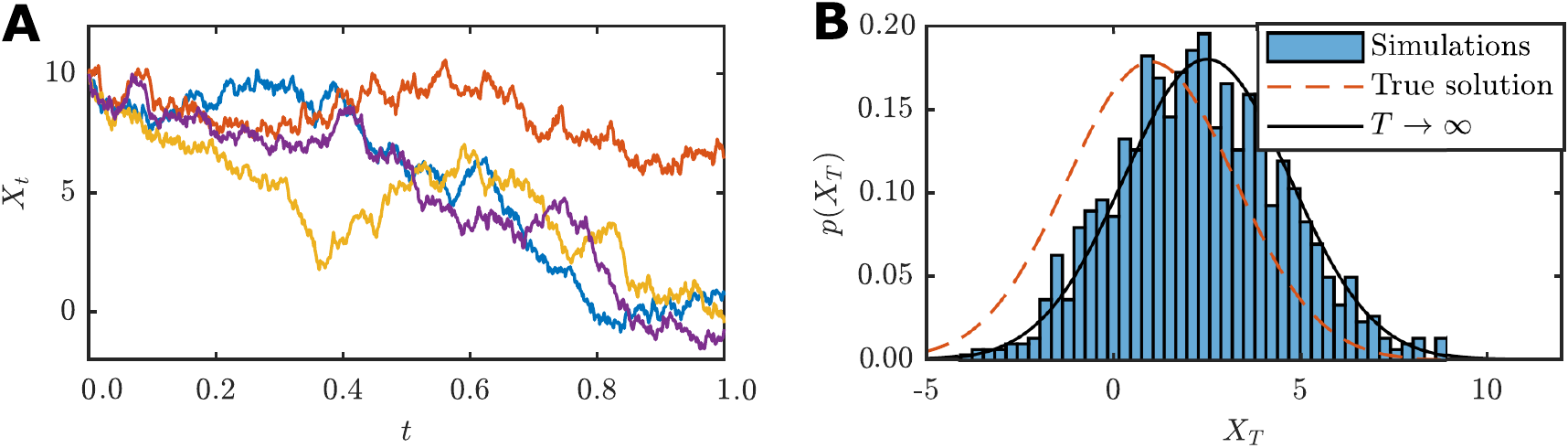}
	\caption{(A) Four example realisations of the Ornstein--Uhlenbeck process with parameters $\mu = 1$, $\gamma = 2$, $\sigma = 2\sqrt{5}$ and initial condition $x_0 = 10$. (B) Empirical distribution of $X_T$ for $T=1$ compared with the exact Fokker--Planck solution and the stationary solution as $T \to \infty$. Simulations are performed using the Euler--Maruyama discretization with time step $\Delta t = 0.001$.}
	\label{fig:ou-sde}
\end{figure}

The Ornstein--Uhlenbeck process is a mean reverting stochastic process with many applications in finance, biology and physics~\citep{Ornstein1930}. The evolution of the continuous state $X_t$ is given by an It\={o} stochastic differential equation (SDE) of the form
\begin{equation}
\begin{split}
\text{d}X_t = \gamma(\mu - X_t)\text{d}t + \sigma \text{d}W_t, \quad\text{if } t > 0, \\
X_0 = x_0, \quad\text{if }t = 0,
\end{split}
\label{eq:ou-sde}
\end{equation}
where $\mu$ is the long-term mean, $\gamma$ is the rate of mean reversion, $\sigma$ is the process volatility, $W_t$ is a Wiener process and $x_0$ is a constant initial condition. Example realisations are shown in Figure~\ref{fig:ou-sde}(A).

We consider data consisting of $N$ independent realisations of the Ornstein--Uhlenbeck processs at time $T < \infty$, that is, $\dat = [X_T^{1},X_T^{2},\ldots, X_T^{N}]$. This inference problem is analytically tractable since the Fokker--Planck equation can be solved to obtain a Gaussian distribution for the data,
\begin{equation}
X_T \sim \mathcal{N}\left(\mu+ (x_0 + \mu)e^{-\gamma t}, 1-\frac{\sigma^2}{2\gamma}e^{-2\gamma t}\right).
\label{eq:ou-fp-sol}
\end{equation}
For demonstration purposes we will assume a solution for the full Fokker--Planck equation is unavailable and perform ABC inference to estimate the volatility parameter, $\sigma$, using stochastic simulation with the Euler--Maruyama  discretisation~\citep{Maruyama1955},
\begin{equation}
X_{t+\Delta t} = X_t + \gamma(\mu - X_t) \Delta t + \sigma \sqrt{\Delta t} \xi_t,
\label{eq:ou-em}
\end{equation}
 where $\xi_t$ is a standard normal variate and $\Delta t$ is a small time step. For the approximate model we take the stationary distribution of the Ornstein--Uhlenbeck process obtained by taking $T \to \infty $ and solving the steady state of the Fokker--Planck equation,  
\begin{equation}
X_\infty \sim \mathcal{N}\left(\mu, \frac{\sigma^2}{2\gamma}\right).
\label{eq:ou-fp-stdy-sol}
\end{equation}
This kind of approximation will often be possible since the steady state Fokker--Planck equation is more likely to be tractable than the transient solution for most SDE models. As shown in Figure~\ref{fig:ou-sde}(B), the stationary solution is a better approximation for the true variance than for the true mean. Therefore, for small $T$ this approximation would be more appropriate as a preconditioner for inference of the volatility parameter, $\sigma$, in \eqref{eq:ou-sde} than for the long-time mean, $\mu$. In general, the performance expected from preconditioning will increase as $T$ increases since the stationary approximation will become more accurate (Supplementary Material).

\begin{figure}[h!]
	\includegraphics[width=\linewidth]{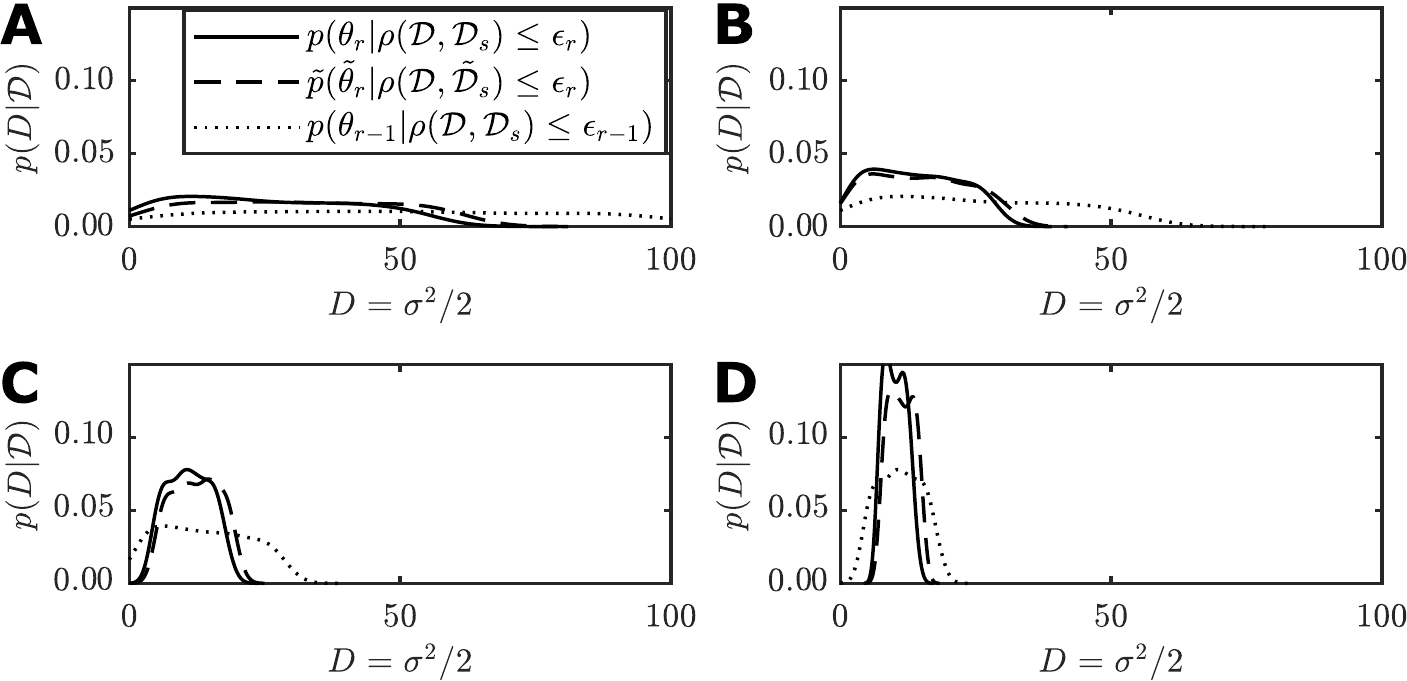}
	\caption{PC-SMC-ABC intermediate steps for the Ornstein-Uhlenbeck SDE example. Each panel demonstrates the transition from threshold $\epsilon_{r-1}$ (dotted line) to $\epsilon_{r}$ (solid line) via the preconditioner (dashed line). (A) $\epsilon_0$ to $\epsilon_1$; (B) $\epsilon_1$ to $\epsilon_2$; (C) $\epsilon_2$ to $\epsilon_3$; and (D) $\epsilon_3$ to $\epsilon_4$. The threshold sequence is $\epsilon_{r} = \epsilon_{r-1}/2$ for $r = 1,2,3,4$ with $\epsilon_ 0 = 6.4$.}
	\label{fig:pc-abc-smc-ou}
\end{figure}

Figure~\ref{fig:pc-abc-smc-ou} demonstrates th results of applying PC-ABC-SMC (Algorithm~\ref{alg:abc_pcsmc}) with $\mathcal{M} = 1000$ particles, showing intermediate distributions, for inferring $D = \sigma^2/2$ given data at $T = 1$, using stochastic simulation for the exact model (\eqref{eq:ou-em}) with $\Delta t = 0.01$, and the stationary distribution (\eqref{eq:ou-fp-stdy-sol}) for the approximate model used in the preconditioning step. 

In this example, all other parameters are treated as known with $x_0 = 10$, $\mu  = 1$ and $\gamma = 2$. Data is generated using $D = 10$ as in Figure~\ref{fig:ou-sde}. In each step, the preconditioner distribution for threshold $\epsilon_{r}$ (dashed line) is a better proposal distribution for the target (solid line) than that of the exact model at threshold $\epsilon_{r-1}$ (dotted line). The overall speedup factor is approximately $1.5$ for this example, and it continues to improve as $T$ increases since \eqref{eq:ou-fp-stdy-sol} becomes an increasingly better approximation to \eqref{eq:ou-fp-sol} (See Supplementary Material). For truly intractable problems, such as the lattice-based random walk models presented in Sections~\ref{sec:allee} and \ref{sec:fkpp}, we obtain superior performance gains of up to a factor of four.

%In reality, these thresholds would normally be chosen to be much smaller to ensure that the posterior mode is closer to the true values.  In this work, we chose modest thresholds to ensure that our results can be obtained in a relatively short period of computational time.   
\FloatBarrier
\subsection{Lattice-based stochastic discrete random walk model}
\label{sec:discrete}
The stochastic discrete model we consider is a lattice-based random walk model that is often used to describe populations of motile cells~\citep{Jin2016b}. The model involves initially placing a population of $N$ agents of size $\delta$ on a lattice, $L$~\citep{Callaghan2006,Simpson2010}, for example an $I \times J$ hexagonal lattice~\citep{Jin2016b}. This hexagonal lattice is defined by a set of indices\\ $L = \{(i,j) : i \in [0,1,\ldots,I-1], j \in [0,1,\ldots,J-1]\}$, and a neighborhood function,
\begin{equation*}
\mathcal{N}(i,j) = \begin{cases}
\{(i-1,j-1),(i,j-1),(i+1,j-1),(i+1,j),(i,j+1),(i-1,j)\} & \text{if } i \text{ is even},\\
\{(i-1,j),(i,j-1),(i+1,j),(i+1,j+1),(i,j+1),(i-1,j+1)\}& \text{if } i \text{ is odd}. 
\end{cases}
%\label{eq:hexNH}
\end{equation*}
Lattice indices are mapped to Cartesian coordinates using
\begin{equation}
(x_i,y_j) = \begin{cases}
\left(i\dfrac{\sqrt{3}}{2}\delta,j\delta\right) & \text{if } i \text{ is even},\\
\left(i\dfrac{\sqrt{3}}{2}\delta, \left(j + \dfrac{1}{2}\right)\delta\right) & \text{if } i \text{ is odd}.
\end{cases}
\end{equation} 
We define an occupancy function such that $C(\ell,t) = 1$ if site $\ell$ is occupied by an agent at time $t \geq 0$, otherwise $C(\ell,t) = 0$. This means that in our discrete model each lattice site can be occupied by, at most, one agent. 

\begin{figure}[h]
	\centering
	\includegraphics[width=0.85\linewidth]{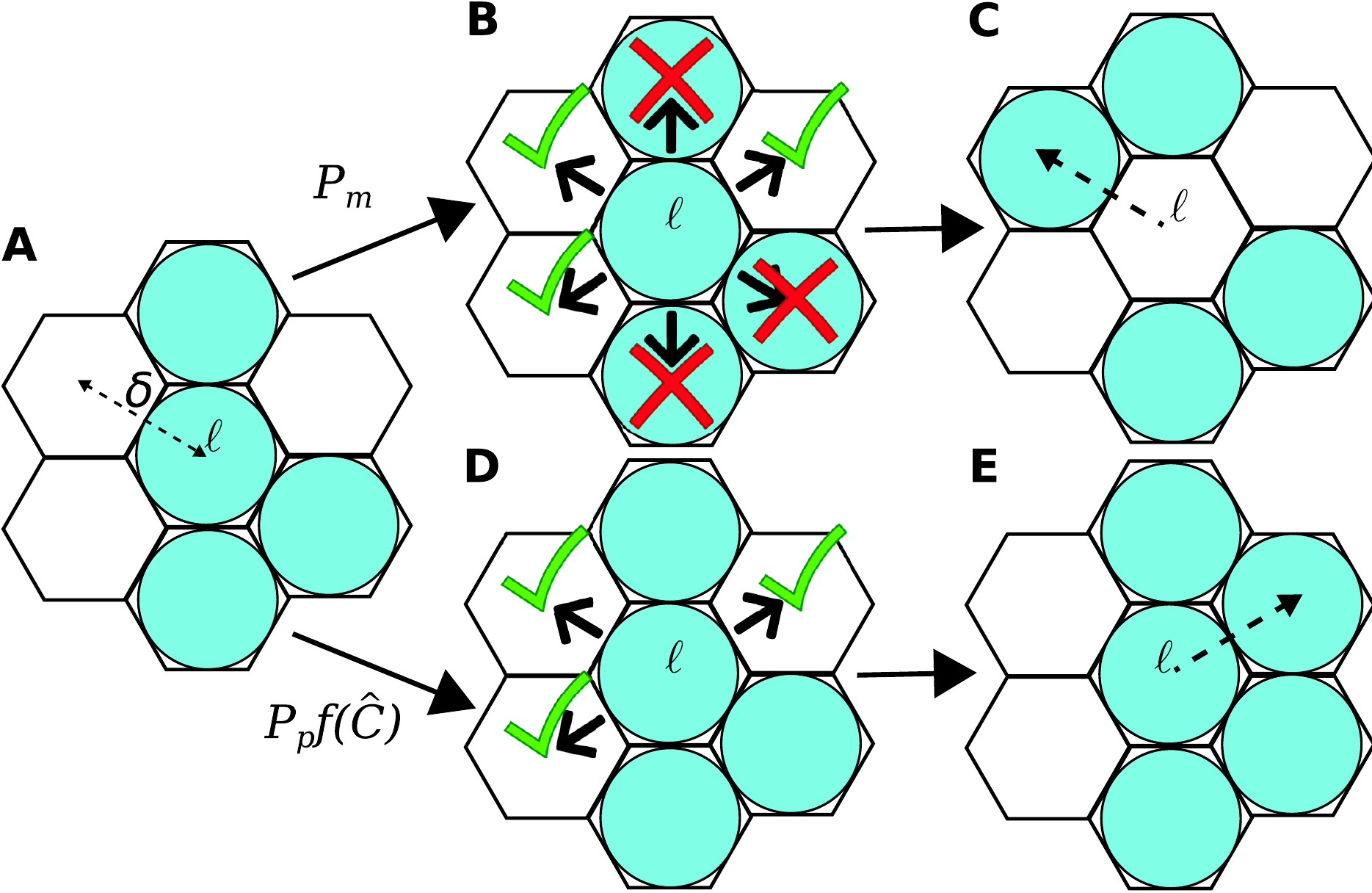}
	\caption{Example of movement and proliferation events in a lattice-based random walk model, using a hexagonal lattice with lattice spacing, $\delta$. (A) An example hexagonal lattice neighborhood $\mathcal{N}(\ell)$. An agent at site $\ell$ attempts a motility event (A)--(C) with probability $P_m$. (B) Motility events are aborted when the randomly selected neighbor site is occupied. (C) The agent moves to the selected site, if unoccupied.  An agent at site $\ell$ attempts a proliferation event (A),(D)--(E) with probability $P_p$. (D) Proliferation events are successful with probability $f(\hat{C}(\ell,t))$, resulting in an unoccupied site being selected. (E) The daughter agent is placed at the selected site and the number of agents in the populations is increased by one.}
	\label{fig:fig1}
\end{figure}

During each discrete time step of duration $\tau$, agents attempt to move with probability $P_m \in [0,1]$ and attempt to proliferate with probability $P_p \in [0,1]$. If an agent at site $\ell$ attempts a motility event, then a neighboring site will be selected uniformly at random. The motility event is aborted if the selected site is occupied, otherwise the agent will move to the selected site  (Figure~\ref{fig:fig1}(A)--(C)). 
For proliferation events, the local neighborhood average occupancy,
\begin{equation*}
\hat{C}(\ell,t) = \frac{1}{6}\sum_{\ell'\in N(\ell)} C(\ell',t),
\end{equation*}
is calculated and a uniform random number $u \sim \mathcal{U}(0,1)$ is drawn. If $u > f(\hat{C}(\ell,t))$, where $f(\hat{C}(\ell,t)) \in [0,1]$ is called the \emph{crowding function}~\citep{Browning2017,Jin2016b}, then the proliferation event is aborted due to local crowding effects and contact inhibition. If $u \leq f(\hat{C}(\ell,t))$, then proliferation is successful and a daughter agent is placed at a randomly chosen unoccupied lattice site in $\mathcal{N}(\ell)$ (Figure~\ref{fig:fig1}(A),(D)--(E)). The evolution of the model is generated through repeating this process though $M$ time steps, $t_1 = \tau,\, t_2 = 2\tau, \ldots,$ $ t_{M} = M \tau$. This approach, based on the work by \citet{Jin2016b}, supports a generic proliferation mechanism since $f(\hat{C}(\ell,t))$ is an arbitrary smooth function satisfying $f(0) = 1$ and $f(K) = 0$, where $K > 0$ is the carrying capacity density. However, in the literature there are also examples that include other mechanisms such as cell-cell adhesion~\citep{Johnston2013}, directed motility~\citep{Binny2016}, and Allee effects~\citep{Bottger2015}. 

%Discrete random walks are stochastic models that enable the microscale behaviours and interactions to be explicitly represented. This allows us to investigate how different local mechanisms and interactions of motility, proliferation and death, affect the global population dynamics. Often, the goal is to use empirical data and statistical inference to elicit the local mechanisms that are essential to real biological or ecological systems~\citep{Ellison2004,Stumpf2014}. 
\FloatBarrier
\subsection{Approximate continuum-limit descriptions}
Discrete models do not generally lend themselves to analytical methods, consequently, their application is intrinsically tied to computationally intensive stochastic simulations and Monte Carlo methods~\citep{Jin2016b}. As a result, it is common practice to approximate mean behavior using differential equations by invoking mean-field assumptions, that is, to treat the occupancy status of lattice sites as independent~\citep{Callaghan2006,Simpson2010}. The resulting approximate continuum-limit descriptions (Supplementary Material) are partial differential equations (PDEs) of the form  
\begin{equation}
\pdydx{\mathcal{C}(x,y,t)}{t} = D\nabla^2\mathcal{C}(x,y,t) + \lambda \mathcal{C}(x,y,t)f(\mathcal{C}(x,y,t)),
\label{eq:cont_PDE}
\end{equation}
where $\mathcal{C}(x,y,t) = \E{C(\ell,t)}$, $D = \lim_{\delta \to 0, \tau \to 0} P_m\delta^2/(4\tau)$ is the diffusivity, $\lambda = \lim_{\tau \to 0} P_p/\tau$ is the proliferation rate with $P_p = \mathcal{O}(\tau)$, and $f(\cdot)$ is the crowding function that is related to the proliferation mechanism implemented in the discrete model~\citep{Browning2017,Jin2016b}. For spatially uniform populations there will be no macroscopic spatial gradients on average, that is $\nabla\mathcal{C}(x,y,t) = \bvec{0}$. Thus, $\mathcal{C}(x,y,t)$ is just a function of time, $\mathcal{C}(t)$, and the continuum limit reduces to an ordinary differential equation (ODE) describing the net population growth,
\begin{equation}
\dydx{\mathcal{C}(t)}{t} = \lambda \mathcal{C}(t)f(\mathcal{C}(t)).
\label{eq:cont_ODE}
\end{equation}
For many standard discrete models, the crowding function is implicitly $f(\mathcal{C}) = 1 - \mathcal{C}$~\citep{Callaghan2006}. That is, the continuum limits in \eqref{eq:cont_PDE} and \eqref{eq:cont_ODE} yield the Fisher-KPP model~\citep{EdelsteinKeshet2005,Murray2002} and the logistic growth model~\citep{Tsoularis2002,Warne2017a}, respectively. However, non-logistic growth models , for example, $f(\mathcal{C}) = (1 - \mathcal{C})^n$ for $n > 1$, have also been considered~\citep{Jin2016b,Simpson2010,Tsoularis2002}.

\subsection{Temporal example: a weak Allee model}
\label{sec:allee}
The Allee effect refers to the reduction in growth rate of a population at low densities. This is particularly well studied in ecology where there are many mechanisms that give rise to this phenomenon~\citep{Taylor2005,Johnston2017}. We introduce an Allee effect into our discrete model by choosing a crowding function of the form
\begin{equation*}
f(\hat{C}(\ell,t)) = \left(1 - \frac{\hat{C}(\ell,t)}{K}\right)\left(\frac{A + \hat{C}(\ell,t)}{K}\right), %\label{eq:wae}
\end{equation*}
where $\hat{C}(\ell,t) \in [0,1]$ is the local density at the lattice site $\ell \in L$, at time $t$, $K > 0$ is the carrying capacity density, and $A$ is the Allee parameter which yields a weak Allee effect for $A \geq 0$~\citep{Wang2019}. Note that smaller values of $A$ entail a more pronounced Allee effect with $A < 0$ leading to a strong Allee effect that can lead to species extinction~\citep{Wang2019}. For simplicity, we only consider the weak Allee effect here, but our methods are general enough to consider any sufficiently smooth $f(\cdot)$. 

Studies in ecology often involve population counts of a particular species over time~\citep{Taylor2005}. In the discrete model, the initial occupancy of each lattice site is independent, and hence there are no macroscopic spatial gradients on average. It is reasonable to summarize simulations of the discrete model at time $t$ by the average occupancy over the entire lattice, $\bar{C}(t) = (1/IJ)\sum_{\ell \in L} C(\ell,t)$.  Therefore, the continuum limit for this case is given by~\citep{Wang2019}
\begin{equation}
\dydx{\mathcal{C}(t)}{t} = \lambda \mathcal{C}(t)\left(1-\frac{\mathcal{C}(t)}{K}\right)\left(\frac{A + \mathcal{C}(t)}{K}\right), \label{eq:cl_allee}
\end{equation}
with $\mathcal{C}(t) = \E{\bar{C}(t)}$, $\lambda = \lim_{\tau \to 0} P_p/\tau$, and $\mathcal{C}(0) = \E{\bar{C}(0)}$.

We generate synthetic time-series ecological data using the discrete model, with observations made at times $t_1 = \tau \times 10^3, t_2= 2\tau \times 10^3,\ldots,t_{10} = \tau \times 10^4$, resulting in data $\dat = [C_{\text{obs}}(t_1),C_{\text{obs}}(t_2), \ldots, C_{\text{obs}}(t_{10})]$ with $C_{\text{obs}}(t) = \bar{C}(t)$ where $\bar{C}(t)$ is the average occupancy at time $t$ for a single realization of the discrete model (Supplementary Material). For this example, we consider an $I \times J$ hexagonal lattice with $I= 80$, $J=68$, and parameters $P_p = 1/1000$, $P_m = 0$, $\delta = \tau = 1$, $K = 5/6$, and $A = 1/10$. Reflecting boundary conditions are applied at all boundaries and a uniform initial condition is applied, specifically, each site is occupied with probability $\Prob{C(\ell,0) = 1} = 1/4$ for all $\ell \in L$, giving $\mathcal{C}(0) = 1/4$.  This combination of parameters is selected since it is known that the continuum limit (\eqref{eq:cl_allee}) will not accurately predict the population growth dynamics of the discrete model in this regime  since $P_p/P_m \gg 1$ (Supplementary Material).

For the inference problem we assume $P_m$ is known, and we seek to compute $\CondPDF{\paramvec}{\dat}$ under the discrete model with $\paramvec = [\lambda, A, K]$ and $\lambda = P_p/\tau$. We utilize uninformative priors, $P_p \sim \mathcal{U}(0,0.005)$, $K \sim \mathcal{U}(0,1)$ and $A \sim \mathcal{U}(0,1)$ with the additional constraint that $A \leq K$,  that is, $A$ and $K$ are not independent in the prior. The discrepancy metric used is the Euclidean distance. For the discrete model, this is 
\begin{equation*}
\discrep{\dat}{\simdat} = \left[\sum_{k=1}^{10} \left(C_{\text{obs}}(t_k) - \bar{C}(t_k)\right)^2\right]^{1/2},
\end{equation*}
where $\bar{C}(t_k)$ is the average occupancy at time $t_k$ of a realization of the discrete model given $\paramvec$. Similarly, for the continuum limit we have
\begin{equation*}
\discrep{\dat}{\approxsimdat} = \left[\sum_{j=1}^{10} \left(C_{\text{obs}}(t_k) - \mathcal{C}(t_k)\right)^2\right]^{1/2},
\end{equation*}
where $\mathcal{C}(t_k)$ is the solution to the continuum limit (\eqref{eq:cl_allee}), computed numerically (Supplementary Material).
We compute the posterior using our PC-SMC-ABC and MM-SMC-ABC methods to compare with SMC-ABC under the continuum limit and SMC-ABC under the discrete model. In each instance, $\mathcal{M} = 1000$ particles are used to approach the target threshold $\epsilon = 0.125$ using the sequence $\epsilon_1,\epsilon_2,\ldots,\epsilon_5$ with $\epsilon_r = \epsilon_{r-1}/2$. In the case of MM-SMC-ABC the tuning parameter is $\alpha = 0.1$. The Gaussian proposal kernels, $\Kernelsub{\paramvec_{r}}{\paramvec_{r-1}}{r}$ and $\approxKernelsub{\paramvec_{r}}{\approxparamvec_{r}}{r}$, are selected adaptively~\citep{Fehlberg1969,Iserles2008} (Supplementary Material).

\begin{figure}[h]
	\centering
	\includegraphics[width=\linewidth]{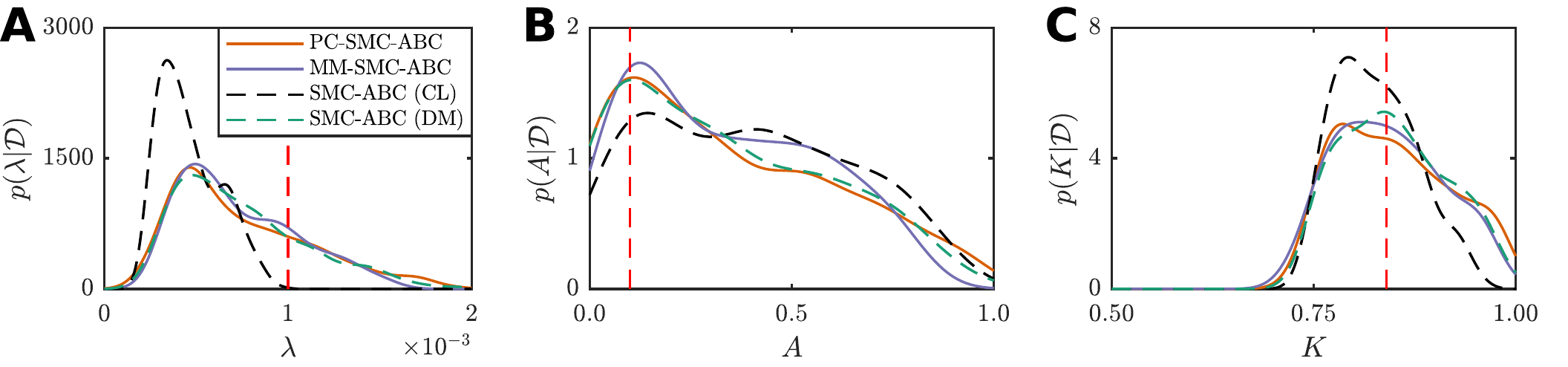}
	\caption{Comparison of estimated posterior marginal densities for the weak Allee model. There is a distinct bias in the SMC-ABC density estimate using the continuum limit (CL) (black dashed) compared with SMC-ABC with the discrete model (DM) (green dashed). However, the density estimates computed using the PC-SMC-ABC (orange solid) and MM-SMC-ABC (purple solid) methods match well with a significantly reduced computational overhead.}
	\label{fig:Allee_results}
\end{figure}

Figure~\ref{fig:Allee_results} and Table~\ref{tab:Allee_results} present the results. SMC-ABC using the continuum-limit model is a poor approximation for SMC-ABC using the discrete model, especially for the proliferation rate parameter, $\lambda$ (Figure~\ref{fig:Allee_results}(a)), which is expected because $P_m = 0$. However, the posteriors estimated using PC-SMC-ABC are an excellent match to the target posteriors estimated using SMC-ABC with the expensive discrete model, yet the PC-SMC-ABC method requires only half the number of stochastic simulations (Table~\ref{tab:Allee_results}). The MM-SMC-ABC method is not quite as accurate as the PC-SMC-ABC method, however, the number of expensive stochastic simulations is reduced by more than a factor of eight (Table~\ref{tab:Allee_results}) leading to considerable increase in computational efficiency. 

\FloatBarrier
\begin{table}[h]
	\caption{Computational performance comparison of the SMC-ABC, PC-SMC-ABC, and MM-SMC-ABC methods for the weak Allee model inference problem. Computations are performed using an Intel$^\text{\textregistered}$ Xeon\texttrademark~E5-2680v3 CPU (2.5 GHz).}
	\begin{tabular}{rcccr}
		Method & Stochastic samples & Continuum samples & Run time (hours) & Speedup \\
		\hline
		SMC-ABC& $28,588$ & $0$ & $47.1$ & $1 \times$ \\
		PC-SMC-ABC & $13,799$ & $58,752$ & $21.1$ & $2\times$\\
		MM-SMC-ABC& $3,342$ & $36,908$  & $5.6$ & $8\times$
	\end{tabular}
	\label{tab:Allee_results}
\end{table}
\FloatBarrier

% two main models, logistic/FKPP and Allee model
\subsection{Spatiotemporal example: a scratch assay}
\label{sec:fkpp}
We now look to a discrete model commonly used in studies of cell motility and proliferation, and use spatially extended data that is typical of \emph{in vitro} cell culture experiments, specifically scratch assays~\citep{Liang2007}. 

In this case we use a crowding function of the form
$f(\hat{C}(\ell,t)) = 1 - {\hat{C}(\ell,t)}/{K}$, where $K > 0$ is the carrying capacity density,
since it will lead to a logistic growth source term in \eqref{eq:cont_PDE} which characterizes the growth dynamics of many cell types~\citep{Simpson2010,Warne2017a}. The discrete model is initialized such that initial density is independent of $y$. Therefore, we summarize the discrete simulation by computing the average occupancy for each $x$ coordinate, that is, we average over the $y$-axis in the hexagonal lattice~\citep{Jin2016b}, that is, $\bar{C}(x,t) = (1/J)\sum_{(x,y) \in L} C((x,y),t)$. Thus, one arrives at the Fisher-KPP model~\citep{EdelsteinKeshet2005,Murray2002} for the continuum limit, 
\begin{equation}
\pdydx{\mathcal{C}(x,t)}{t} = D \pddydx{\mathcal{C}(x,t)}{x} + \lambda \mathcal{C}(x,t)\left(1-\frac{\mathcal{C}(x,t)}{K}\right),
\label{eq:cont_FKPP}
\end{equation}
where $\mathcal{C}(x,t) = \E{\bar{C}(x,t)}$, $D = \lim_{\delta \to 0, \tau \to 0} P_m\delta^2/(4\tau)$, and $\lambda = \lim_{\tau \to 0} P_p/\tau$.

Just as with the weak Allee model, here we generate synthetic spatiotemporal cell culture data using the discrete model. Observations are made at times $t_1 = 3\tau \times 10^2,$ $t_2= 6\tau \times 10^2,\ldots,t_{10} = 3\tau \times 10^3$, resulting in data 
\begin{equation*}
\dat = \left[\begin{matrix}
C_{\text{obs}}(x_1,t_1) & C_{\text{obs}}(x_1,t_2) & \ldots & C_{\text{obs}}(x_1,t_{10}) \\
C_{\text{obs}}(x_2,t_1) & C_{\text{obs}}(x_2,t_2) & \ldots & C_{\text{obs}}(x_2,t_{10}) \\
\vdots & \vdots & \ddots & \vdots \\
C_{\text{obs}}(x_I,t_1) & C_{\text{obs}}(x_I,t_2) & \ldots & C_{\text{obs}}(x_I,t_{10}) 
\end{matrix}\right],
\end{equation*}
with $C_{\text{obs}}(x,t) = \bar{C}(x,t)$ where $\bar{C}(x,t)$ is the average occupancy over sites $(x,y_1), (x,y_2),$ $\ldots,(x,y_J)$ at time $t$ for a single realization of the discrete model. As with the weak Allee model, we consider an $I \times J$ hexagonal lattice with $I=80$, $J=68$, and parameters $P_p = 1/1000$, $P_m = 1$, $\delta = \tau = 1$ and $K = 5/6$. We simulate a scratch assay by specifying the center 20 cell columns ($31 \leq i \leq 50$) to be initially unoccupied, and apply a uniform initial condition outside the scratch area such that $\E{C(\ell,0)} = 1/4$ overall. Reflecting boundary conditions are applied at all boundaries. Note, we have selected a parameter regime with $P_p/P_m \ll 1$ for which the continuum limit is an accurate representation of the discrete model average behavior (Supplementary Material).  

Since we have spatial information for this problem, we assume $P_m$ is also an unknown parameter and perform inference on the discrete model to compute $\CondPDF{\paramvec}{\dat}$ with\\ $\paramvec = [\lambda, D, K]$, $\lambda = P_p/\tau$, and $D = P_m\delta^2/4\tau$. We utilize uninformative priors,\\ $P_p \sim \mathcal{U}(0,0.008)$, $P_m \sim \mathcal{U}(0,1)$, and $K \sim \mathcal{U}(0,1)$. For the discrepancy metric we use the Frobenius norm; for the discrete model,  this is 
\begin{equation*}
\discrep{\dat}{\simdat} = \left[\sum_{k=1}^{10}\sum_{i=1}^I\left(C_{\text{obs}}(x_i,t_k) - \bar{C}(x_i,t_k)\right)^2\right]^{1/2},
\end{equation*}
where $\bar{C}(x_i,t_k)$ is the average occupancy at site $x_i$ at time $t_k$ of a realization of the discrete model given parameters $\paramvec$. Similarly, for the continuum limit we have
\begin{equation*}
\discrep{\dat}{\approxsimdat} = \left[\sum_{k=1}^{10}\sum_{i=1}^I\left(C_{\text{obs}}(x_i,t_k) - \mathcal{C}(x_i,t_k)\right)^2\right]^{1/2},
\end{equation*}
where $\mathcal{C}(x_i,t_k)$ is the solution to the continuum-limit PDE (\eqref{eq:cont_FKPP}), computed using a backward-time, centered-space finite difference scheme with fixed-point iteration and adaptive time steps~\citep{Simpson2007,Sloan1999} (Supplementary Material). 
We estimate the posterior using our PC-SMC-ABC and MM-SMC-ABC methods to compare with SMC-ABC using the continuum limit and SMC-ABC using the discrete model. In each case, $\mathcal{M} = 1000$ particles are used to approach the target threshold, $\epsilon = 2$, using the sequence $\epsilon_1,\epsilon_2,\ldots,\epsilon_5$ with $\epsilon_r = \epsilon_{r-1}/2$. In the case of MM-SMC-ABC the tuning parameter is $\alpha = 0.1$. Again, Gaussian proposal kernels, $\Kernelsub{\paramvec_{r}}{\paramvec_{r-1}}{r}$ and $\approxKernelsub{\paramvec_{r}}{\approxparamvec_{r}}{r}$, are selected adaptively (Supplementary Material).

\begin{figure}[h]
	\centering
	\includegraphics[width=\linewidth]{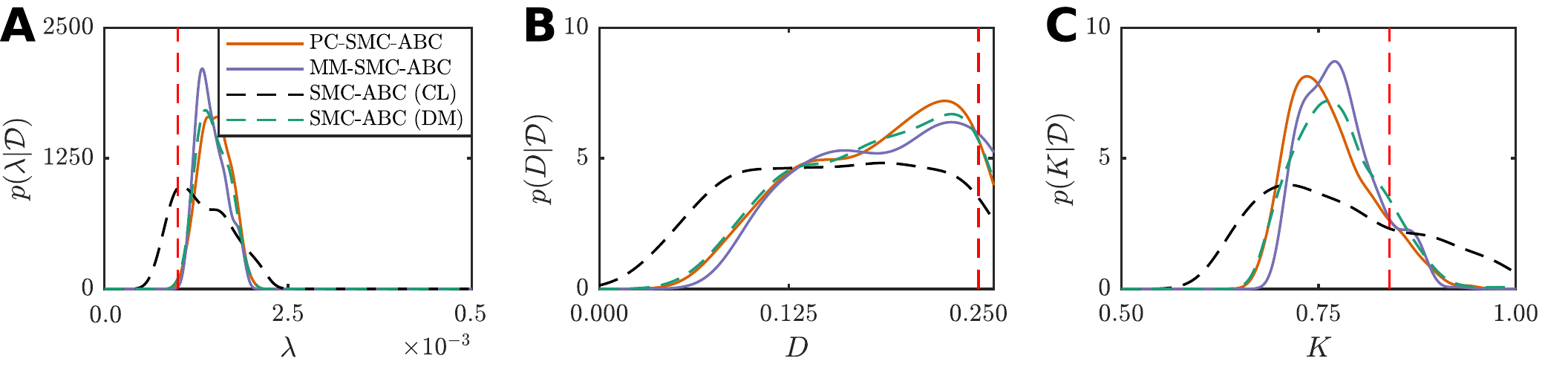}
	\caption{Comparison of estimated posterior marginal densities for the scratch assay model. There is a distinct bias in the SMC-ABC density estimate using the continuum limit (CL) (black dashed) compared with SMC-ABC with the discrete model (DM) (green dashed). However, the density estimates computed using the PC-SMC-ABC (orange solid) and MM-SMC-ABC (purple solid) methods match well with a reduced computational overhead.}
	\label{fig:FKPP_results}
\end{figure}

Results are shown in Figure~\ref{fig:FKPP_results} and Table~\ref{tab:FKPP_results}. Despite the continuum limit being a good approximation of the discrete model average behavior, using solely this continuum limit in the inference problem still leads to bias. Just as with the weak Allee model, both PC-SMC-ABC and MM-SMC-ABC methods produce a more accurate estimate of the SMC-ABC posterior density with the discrete model. Overall, PC-SMC-ABC is unbiased, however, MM-SMC-ABC is still very accurate. The main point for our work is that the PC-SMC-ABC and MM-SMC-ABC methods both produce posteriors that are accurate compared with the expensive stochastic inference problem, whereas the approximate model alone does not. From Table~\ref{tab:FKPP_results}, both PC-SMC-ABC and MM-SMC-ABC require a reduced number of stochastic simulations of the discrete model compared with direct SMC-ABC. For PC-SMC-ABC, the reduction is almost a factor of four and, for MM-SMC-ABC, the reduction is almost a factor of eleven.  

\FloatBarrier
\begin{table}[h]
	\caption{Computational performance comparison of the SMC-ABC, PC-SMC-ABC, and MM-SMC-ABC methods, using the scratch assay model inference problem. Computations are performed using an Intel$^\text{\textregistered}$ Xeon\texttrademark~E5-2680v3 CPU (2.5 GHz).}
	\begin{tabular}{rcccr}
		Method & Stochastic samples & Continuum samples & Run time (hours)  & Speedup \\
		\hline
		SMC-ABC& $46,435$ & $0$ & $20.6$ & $1\times$\\
		PC-SMC-ABC & $13,949$ & $13,179$ & $5.6$ & $4\times$\\
		MM-SMC-ABC& $4,457$ & $10,594$  & $1.9$ & $11\times$\\
	\end{tabular}
	\label{tab:FKPP_results}
\end{table}
\FloatBarrier

\subsection{A guide to selection of $\alpha$ for MM-SMC-ABC}
\label{sec:selalpha}
The performance of MM-SMC-ABC is dependent on the tuning parameter $\alpha \in [0,1]$. Since MM-SMC-ABC will only propagate $\ceil{\alpha\mathcal{M}}$ particles based on the expensive stochastic model, $\alpha$ can be considered as a target computational cost reduction factor with $1/\alpha$ being the target speed up factor. However, intuitively there will be a limit as to how small one can choose $\alpha$ before the statistical error incurred from the estimates of $\boldsymbol{\mu}_r$ and $\boldsymbol{\Sigma}_r$ is large enough to render the approximate moment matching transform inaccurate. 
It is non-trivial to analyze MM-SMC-ABC to obtain a theoretical guideline for choosing $\alpha$, therefore we perform a computational benchmark to obtain a heuristic. It should be noted that we use an SMC-ABC sampler with $\mathcal{M} = 1000$ as a benchmark for accuracy and performance. As a result, it may be that repeating the analysis with larger $\mathcal{M}$ could lead to a smaller optimal $\alpha$.%  \todo{Highlight that larger $\mathcal{M}$ could enable smaller $\alpha$} 

Here, using different values for $\alpha$ we repeatedly solve the weak Allee model (Section~\ref{sec:allee}) and the scratch assay model (Section~\ref{sec:fkpp}). For both inverse problems we applied MM-SMC-ABC under identical conditions as in Sections~\ref{sec:allee} and \ref{sec:fkpp} with the exception of the tuning parameter $\alpha$ that takes values from the sequence $\left\{\alpha_k\right\}_{k=0}^5$ with $\alpha_0 = 0.8$ and $\alpha_k = \alpha_{k-1}/2$ for $k > 0$. For each $\alpha_k$ in the sequence, we consider $N$ independent applications of MM-SMC-ABC. The computational cost for each $\alpha_k$ is denoted by $\text{Cost}(\alpha_k)$ and represents the run time in seconds for an application of MM-SMC-ABC with tuning parameter $\alpha_k$. We also calculate an error metric,
\begin{equation*}
\text{Error}(\alpha_k) = \mathcal{E}\left(\paramspace_R,\paramspace_R(\alpha_k),P\right),
\end{equation*}
where $\paramspace_R = \left\{\paramvec_R^i\right\}_{i=1}^{\mathcal{M}}$ is a set of particles from an application of SMC-ABC using the expensive stochastic model, and $\paramspace_R(\alpha_k) = \left(\left\{\paramvec_R^i\right\}_{i=1}^{\ceil{\alpha_k\mathcal{M}}} \cup \left\{\bar{\paramvec}_R^i\right\}_{i=1}^{\floor{(1-\alpha_k)\mathcal{M}}}\right)$ is the pooled exact and approximate transformed particles from the $j$th application of MM-SMC-ABC. For $P \in \mathbb{N}$, the function $ \mathcal{E}(\cdot,\cdot,P)$ is the $P$-th order empirical moment-matching distance function ~\citep{Liao2015,Lillacci2010,Zechner2012}, given by
\begin{equation*}
\mathcal{E}(\bvec{X},\bvec{Y},P) = \sum_{m=0}^P \sum_{\bvec{b} \in S_m} \frac{1}{|S_m|^m} \left(\frac{\hat{\mu}(\bvec{X})^\bvec{b} - \hat{\mu}(\bvec{Y})^\bvec{b}}{\hat{\mu}(\bvec{X})^\bvec{b}}\right)^2,
\end{equation*} 
for two sample sets $\bvec{X} = \left\{\bvec{x}_1,\bvec{x}_2, \ldots, \bvec{x}_{\mathcal{M}} \right\}$ and $\bvec{Y} = \left\{\bvec{y}_1,\bvec{y}_2, \ldots, \bvec{y}_{\mathcal{M}} \right\}$  with $\bvec{x}_i,\bvec{y}_i \in \mathbb{R}^n$ for $n \geq 1$, and $S_m = \left\{\bvec{b} : \bvec{b} \in \mathbb{N}^n, \|\bvec{b}\| = m \right\} $. For any $n$-dimensional discrete vector $\bvec{b} = \left[b_1,b_2,\ldots,b_n\right]^{\text{T}}\in \mathbb{N}^n$, then $\hat{\mu}(\bvec{X})^\bvec{b}$ is the $\bvec{b}$th empirical raw moment of the sample\\ set $\bvec{X}$,
\begin{equation*}
\hat{\mu}(\bvec{X})^\bvec{b} = \frac{1}{\mathcal{M}}\sum_{i=1}^{\mathcal{M}} \bvec{x}_i^{\bvec{b}},
\end{equation*}
where $\bvec{x}_i^{\bvec{b}} = x_{i,1}^{b_1}\times x_{i,2}^{b_2} \times \cdots \times x_{i,n}^{b_n}$. Note that $P$ must be greater than the number of moments that are matched in the approximate transform~(\eqref{eq:mmtfapprox}) to ensure that MM-SMC-ABC is improving the accuracy in higher moments also.

\begin{figure}[h]
	\centering
	\includegraphics[width=\textwidth]{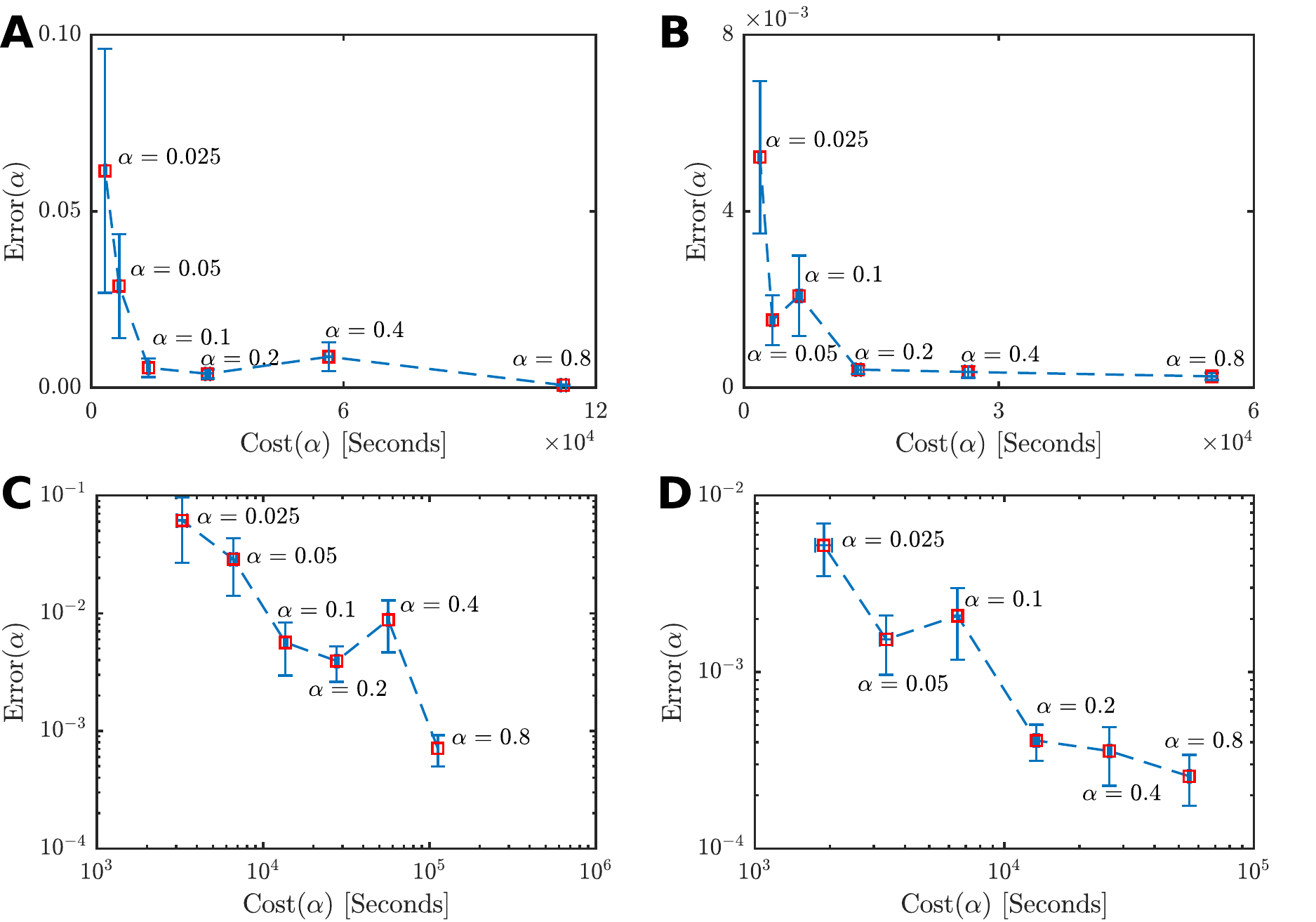}
	\caption{Error versus cost plots for different values of the tuning parameter $\alpha$. Averages and standard errors are shown for $N = 10$ independent applications of the MM-SMC-ABC method to the (A) weak Allee effect model and (B) the scratch assay model. Results in (A)--(B) are shown in (C)--(D) in a log scale for clarity.}
	\label{fig:costvsmomenterror}
\end{figure}

We estimate the average $\text{Cost}(\alpha_k)$ and $\text{Error}(\alpha_k)$ for each value of $\alpha_k$ for both the weak Allee effect and the scratch assay inverse problems. Figure~\ref{fig:costvsmomenterror} displays the estimates and standard errors given $P=6$ and $N = 10$, with the value of $\alpha_k$ shown. We emphasize that $P > 2$, that is our error measure compares the first six moments while only two moments are matched.
There is clearly a threshold for $\alpha$, below which the error becomes highly variable. For both the weak Allee effect model (Figure~\ref{fig:costvsmomenterror}(A),(C)) and the scratch assay model (Figure~\ref{fig:costvsmomenterror}(B),(D)), the optimal choice of $\alpha$ is located between $\alpha_2 = 0.2$ and $\alpha_4 = 0.05$. Therefore, we suggest a heuristic of $\alpha \in [0.1,0.2]$ to be a reliable choice. If extra performance is needed $\alpha \in [0.05, 0.1)$ may also be acceptable, but if accuracy is of the utmost importance then $\alpha \approx 0.2$ seems to be the most robust choice. This experiment also provides insight in to the consistency and stability of the MM-SMC-ABC method, where $\alpha \geq 0.1$ leads to results that are consistently fast and have low variability in the error metric. While further work is required to assess theoretically the stability and consistency properties of this method, these numerical results are promising. In general, the choice of optimal $\alpha$ is still an open problem and is likely to be impacted by the specific nature of the relationship between the exact model and the approximate model.

\FloatBarrier

\subsection{Summary}
This section presented numerical examples to demonstrate our new methods, PC-SMC-ABC and MM-SMC-ABC, for ABC inference with expensive stochastic discrete models. The tractable Ornstein--Uhlenbeck process was used to highlight the mechanisms leading to the performance improvements of PC-ABC-SMC. Then two examples based on lattice-based random walks were used to demonstrate the efficacy of both PC-SMC-ABC and MM-SMC-AB. In the weak Allee model example, data were generated using parameters that violate standard continuum-limit assumptions; in the scratch assay model example, the Fisher-KPP continuum limit is known to be a good approximation in the parameter regime of the generated data. In both examples, final inferences are biased when the continuum limit is exclusively relied on in the SMC-ABC sampler. However, the results from our new algorithms, PC-SMC-ABC and MM-SMC-ABC, show significantly more accurate posteriors can be computed at a fraction of the cost of the full SMC-ABC using the discrete model, with speed improvements over an order of magnitude.

As mentioned in Section~\ref{sec:mmsmcabc}, the tuning parameter, $\alpha$, in the MM-SMC-ABC method effectively determines the trade-off between the computational speed of the approximate model and the accuracy of the expensive stochastic model. The values $\alpha = 0$ and $\alpha = 1$ correspond to performing inference exclusively with, respectively, the continuum limit and the stochastic discrete model. Based on numerical experimentation, we find that $\alpha \approx 0.1$ is quite reasonable, however, this conclusion will be dependent on the specific model, the parameter space dimensionality, and the number of particles used for the SMC scheme.

\section{Discussion}
\label{sec:discuss}

In the life sciences, computationally challenging stochastic discrete models are routinely used to characterize the dynamics of biological populations~\citep{Codling2008,Callaghan2006,Simpson2010}. In practice, approximations such as the mean-field continuum limit are often derived and used in place of the discrete model for analysis and, more recently, for inference. However, parameter inferences will be biased when the approximate model is solely utilized for inference, even in cases when the approximate model provides an accurate description of the average behaviour of the stochastic model.

We provide a new approach to inference for stochastic models that maintains all the methodological benefits of working with discrete mathematical models, while avoiding the computational bottlenecks of relying solely upon repeated expensive stochastic simulations. Our two new algorithms, PC-SMC-ABC and MM-SMC-ABC, utilize samples from the approximate model inference problem in different ways to accelerate SMC-ABC sampling. The PC-SMC-ABC method is asymptotically unbiased, and we demonstrate computational improvements of up to a factor of almost four are possible. While potentially biased, MM-SMC-ABC can provide further improvements. In general, the expected speedup is around $1/\alpha$, and $\alpha \approx 0.1$ is reasonable based on our numerical investigations. For larger values of $\mathcal{M}$ it may be that even smaller values of $\alpha$ could be effective. %\todo{expand on choice of $\mathcal{M}$ and $\alpha$}

There are some assumptions in our approach that could be generalized in future work. First, in PC-SMC-ABC, we assume that the condition in~\eqref{eq:effcond} holds for all $\epsilon_r$; this is reasonable for the models we consider since we never observe a decrease in performance. However, it may be possible for the bias in the approximate model to be so extreme for some $\epsilon_r$ that the condition in~\eqref{eq:effcond} is violated, leading to a decrease in performance at specific generations. Acceptance probabilities could be estimated by performing a small set of trial samples from both $\eta_{{r-1}}(\paramvec_{r-1})$ and $\tilde{\eta}_r(\paramvec_r)$ proposal mechanisms, enabling automatic selection of the optimal proposal mechanism. Second, 
in the moment matching transform proposed in~\eqref{eq:mmtf}, we use two moments only as this is sufficient for the problems we consider here with numerical examples demonstrating accuracy in the first six moments. However, our methodology is sufficiently flexible that additional moments can be incorporated if necessary. While including higher moments will improve the accuracy of the moment-matching transform, more samples from the exact model will be required to achieve initial estimates of these moments resulting in eroded performance.% due to a higher optimal value of $\alpha$.  
%The moment matching transform proposed in~\eqref{eq:mmtf} will not always be valid. For example, if the number of modes present in the posterior under the continuum limit differs to the number of modes in the posterior under the discrete model, then the transform will not be able to capture this exactly, that is, the transform continuum limit samples are a poor emulator to the posterior under the discrete model. Therefore, it may be necessary in some cases employ a more general moment matching transform that includes higher order moments. 

%Lastly, both methods will be more effective when the bias from the approximate inference problem is low, since inference on the approximate model is computationally very cheap; testing the bias of the approximate inference problem will be difficult to test \emph{a priori}.   

While the performance improvements we demonstrate here are significant, it is also possible to obtain improvements of similar order through detailed code optimization techniques  applied to standard SMC-ABC. We emphasize that our schemes would also benefit from such optimizations as advanced vectorization and parallelization to further improve their performance~\citep{Lee2010,Warne2019c}. Our algorithm extensions are also more direct to implement over advanced high performance computing techniques for acceleration of computational schemes.

There are many extensions to our methods that could be considered. We have based our presentation on a form of an SMC-ABC sampler that uses a fixed sequence of thresholds. However, the ideas of using the preconditioning distribution, as in PC-SMC-ABC, and the moment matching transform, as in MM-SMC-ABC, are applicable to SMC schemes that adaptively select thresholds~\citep{Drovandi2011}. 
Recently, there have been a number of state-of-the-art inference schemes introduced based on multilevel Monte Carlo (MLMC)~\citep{Giles2015,Warne2019b}. Our new SMC-ABC schemes could exploit MLMC to combine samples from all acceptance thresholds using a coupling scheme and bias correction telescoping summation, such as in the work of \cite{Jasra2019} or \cite{Warne2018}. Early accept/rejection schemes, such as those considered by \citet{Prangle2016}, \citet{Prescott2020}, and \citet{Lester2018}, could also be introduced for the sampling steps involving the expensive discrete model. Lastly, the PC-SMC-ABC and the MM-SMC-ABC methods could also be applied together and possibly lead to a compounding effect in the performance. 
Delayed acceptance schemes~\citep{Banterle2019,Everitt2020,Golightly2015} are also an alternative approach with similar motivations to the methods we propose in this work. However, these approaches can be highly sensitive to false negatives, that is, cases where a particular value of $\paramvec$ would be rejected under the approximate model but accepted under the exact model . Our PC-SMC-ABC approach is not affected by false negatives due to the use of the second set of proposal kernels. %Due to this sensitivity to false negatives, the delayed acceptance form of ABC can be biased. Our PC-SMC-ABC approach is not affected by false negatives due to the use of the second set of proposal kernels, $\approxKernelsub{\paramvec}{\approxparamvec}{r}$. This ensures that PC-SMC-ABC is unbiased, which is a distinct advantage over delayed acceptance ABC.  

We have demonstrated our methods using a two-dimensional lattice-based discrete random-walk model that leads to mean-field continuum-limit approximations with linear diffusion and a source term of the form $\lambda \mathcal{C}f(\mathcal{C})$. However, our methods are more widely applicable. We could further generalize the model to deal with a more general class of reaction-diffusion continuum limits involving nonlinear diffusion~\citep{Warne2019,Witelski1995} and generalized proliferation mechanisms~\citep{Simpson2013, Tsoularis2002}. Our framework is also relevant to lattice-free discrete models~\citep{Codling2008,Browning2018} and higher dimensional lattice-based models~\citep{Browning2019}; we expect the computational improvements will be even more significant in this case. Many other forms of model combinations are also be possible. For example, a sequence of continuum models of increasing complexity could be considered, as in \cite{Browning2019}. Alternatively, a sequence of numerical approximations of increasing accuracy could be used for inference using a complex target PDE model~\citep{Cotter2010}. Linear mapping approximations of higher order chemical reaction network models, such as in \citet{Cao2018}, could also exploit our approach. Another relevant and very general application in systems biology is utilize reaction rate equations, that are deterministic ODEs, as approximations to stochastic chemical kinetics models~\citep{Higham2008,Wilkinson2009}. 

Of course, not all approximate models will necessarily provide performance improvements. As demonstrated for the Ornstein--Uhlenbeck example (Section~\ref{sec:ousde}), the stationary distribution will be more appropriate for inference of $\sigma$ rather than $\mu$ with the approximation improving as the model sample time $T$ increases. However, as shown for lattice-based random walk models (Sections~\ref{sec:allee} and \ref{sec:fkpp}), even when the assumptions associated with the approximation do not hold, it is still possible to improve sampling with PC-SMC-ABC and MM-ABC-SMC. Therefore, we suggest that approximations that are derived from some limiting, averaged behavior of the exact model will be good initial candidates for our methods. Semi-automated model reduction techniques also are potential approaches to obtain approximations~\citep{Transtrum2014} that could be investigated in the future. % \todo{Mention general model reductions such as MBAM and discuss guides on selection of an approximate model (e.g., derived from the exact model like mean-fields or steady states)}.  

In this work, novel methods have been presented for exploiting approximate models to accelerate Bayesian inference for expensive stochastic models. We have shown that, even when the approximation leads to biased parameter inferences, it can still inform the proposal mechanisms for ABC samplers using the stochastic model.  Our numerical examples show performance improvements of more than tenfold. These substantial computational improvements  are promising and expands the feasibility of Bayesian analysis for problems involving expensive stochastic models.

\paragraph{Software availability}
Numerical examples presented in this work are available from GitHub\\ \href{https://github.com/ProfMJSimpson/Warne_RapidBayesianInference_2019}{https://github.com/ProfMJSimpson/Warne\_RapidBayesianInference\_2019}.
\paragraph{Acknowledgements}
This work was supported by the Australian Research Council (DP170100474). D.J.W. and M.J.S. acknowledge continued support from the Centre for Data Science at the Queensland University of Technology.  D.J.W. and M.J.S. are members of the Australian Research Council Centre of Excellence for Mathematical and Statistical Frontiers.	R.E.B. would like to thank the Leverhulme Trust for a Leverhulme Research Fellowship, the Royal Society for a Wolfson Research Merit Award, and the BBSRC for funding via BB/R00816/1. M.J.S. appreciates support from the University of Canterbury Erskine Fellowship.
Computational resources where provided by the eResearch Office, Queensland University of Technology. The authors thank Wang Jin for helpful discussions.

\newpage
\begin{appendices}
	\appendix

	\section{Analysis of PC-SMC-ABC}
	\numberwithin{equation}{section}
	\numberwithin{figure}{section}
	\numberwithin{algorithm}{section}
	\numberwithin{table}{section}
	\setcounter{equation}{0}
	
	Here we demonstrate that the PC-SMC-ABC method is asymptotically unbiased. Using similar arguments to Del~Moral et al.~\cite{DelMoral2006} and Sisson et al.~\cite{Sisson2007}, we show that the weighting update scheme can be interpreted as importance sampling on the joint space  distribution of particle trajectories and that this joint target density admits the target posterior density as a marginal. 
	
	Given the target sequence, $\{\CondPDF{\paramvec_r}{\discrep{\dat}{\simdat} \leq \epsilon_r}\}_{r=1}^R$, and approximate sequence, $\{\approxCondPDF{\approxparamvec_r}{\discrep{\dat}{\approxsimdat} \leq \epsilon_r}\}_{r=1}^R$, with prior $\PDF{\paramvec_0}$, $\epsilon_{r-1} > \epsilon_r$ and proposal kernels $\approxKernelsub{\paramvec_{r}}{\approxparamvec_{r}}{r}$ and $\Kernelsub{\approxparamvec_{r}}{\paramvec_{r-1}}{r}$, we write the unnormalized weighting update scheme for PC-SMC-ABC. That is,
	\begin{equation}
	\tilde{w}_r^i = w_{r-1}^i \frac{\CondPDF{\approxparamvec_r^i}{\discrep{\dat}{\approxsimdat} \leq \epsilon_r}\BKernel{\paramvec_{r-1}^i}{\approxparamvec_{r}^i}{r-1}}{\CondPDF{\paramvec_{r-1}^i}{\discrep{\dat}{\simdat} \leq \epsilon_{r-1}}\Kernelsub{\approxparamvec_{r}^i}{\paramvec_{r-1}^i}{r}},
	\label{eq:approxw}
	\end{equation}
	and
	\begin{equation}
	w_r^i = \tilde{w}_{r}^i \frac{\CondPDF{\paramvec_r^i}{\discrep{\dat}{\simdat} \leq \epsilon_r}\approxBKernel{\approxparamvec_{r}^i}{\paramvec_{r}^i}{r}}{\CondPDF{\approxparamvec_{r}^i}{\discrep{\dat}{\approxsimdat} \leq \epsilon_{r}}\approxKernelsub{\paramvec_{r}^i}{\approxparamvec_{r}^i}{r}},
	\label{eq:exactw}
	\end{equation}
	where $\BKernel{\paramvec_{r-1}^i}{\approxparamvec_{r}^i}{r-1}$ and $\approxBKernel{\approxparamvec_{r}^i}{\paramvec_{r}^i}{r}$ are arbitrary backwards kernels. Note that Algorithm~2 in the main manuscript is not expressed in terms of these arbitrary kernels, but rather we utilize optimal backwards kernels. To proceed we substitute \eqref{eq:approxw} into \eqref{eq:exactw} and simplify as follows,
	\begin{align}
	w_r^i &= w_{r-1}^i \frac{\CondPDF{\paramvec_r^i}{\discrep{\dat}{\simdat} \leq \epsilon_r}\approxBKernel{\approxparamvec_{r}^i}{\paramvec_{r}^i}{r}}{\CondPDF{\approxparamvec_{r}^i}{\discrep{\dat}{\approxsimdat} \leq \epsilon_{r}}\approxKernelsub{\paramvec_{r}^i}{\approxparamvec_{r}^i}{r}} \times \frac{\CondPDF{\approxparamvec_r^i}{\discrep{\dat}{\approxsimdat} \leq \epsilon_r}\BKernel{\paramvec_{r-1}^i}{\approxparamvec_{r}^i}{r-1}}{\CondPDF{\paramvec_{r-1}^i}{\discrep{\dat}{\simdat} \leq \epsilon_{r-1}}\Kernelsub{\approxparamvec_{r}^i}{\paramvec_{r-1}^i}{r}} \notag \\
	&= w_{r-1}^i \frac{\CondPDF{\paramvec_r^i}{\discrep{\dat}{\simdat} \leq \epsilon_r}\approxBKernel{\approxparamvec_{r}^i}{\paramvec_{r}^i}{r}\BKernel{\paramvec_{r-1}^i}{\approxparamvec_{r}^i}{r-1}}{\CondPDF{\paramvec_{r-1}^i}{\discrep{\dat}{\simdat} \leq \epsilon_{r-1}}\approxKernelsub{\paramvec_{r}^i}{\approxparamvec_{r}^i}{r}\Kernel{\approxparamvec_{r}^i}{\paramvec_{r-1}^i}{r}} \times \frac{\CondPDF{\approxparamvec_r^i}{\discrep{\dat}{\approxsimdat} \leq \epsilon_r}}{\CondPDF{\approxparamvec_r^i}{\discrep{\dat}{\approxsimdat} \leq \epsilon_r}} \notag \\
	&= w_{r-1}^i \frac{\CondPDF{\paramvec_r^i}{\discrep{\dat}{\simdat} \leq \epsilon_r}\approxBKernel{\approxparamvec_{r}^i}{\paramvec_{r}^i}{r}\BKernel{\paramvec_{r-1}^i}{\approxparamvec_{r}^i}{r-1}}{\CondPDF{\paramvec_{r-1}^i}{\discrep{\dat}{\simdat} \leq \epsilon_{r-1}}\approxKernelsub{\paramvec_{r}^i}{\approxparamvec_{r}^i}{r}\Kernelsub{\approxparamvec_{r}^i}{\paramvec_{r-1}^i}{r}}.
	\label{eq:simup}
	\end{align}
	Now, recursively expand the weight update sequence (\eqref{eq:simup}) to obtain the final weight for the $i$th particle,
	\begin{align}
	w_R^i &= \frac{\CondPDF{\paramvec_1^i}{\discrep{\dat}{\simdat} \leq \epsilon_1}\approxBKernel{\approxparamvec_{1}^i}{\paramvec_{1}^i}{1}\BKernel{\paramvec_{0}^i}{\approxparamvec_{1}^i}{0}}{\PDF{\paramvec_{0}^i}\approxKernelsub{\paramvec_{1}^i}{\approxparamvec_{1}^i}{1}\Kernelsub{\approxparamvec_{1}^i}{\paramvec_{0}^i}{1}} \notag \\  &\quad \times \prod_{r=2}^R \frac{\CondPDF{\paramvec_r^i}{\discrep{\dat}{\simdat} \leq \epsilon_r}\approxBKernel{\approxparamvec_{r}^i}{\paramvec_{r}^i}{r}\BKernel{\paramvec_{r-1}^i}{\approxparamvec_{r}^i}{r-1}}{\CondPDF{\paramvec_{r-1}^i}{\discrep{\dat}{\simdat} \leq \epsilon_{r-1}}\approxKernelsub{\paramvec_{r}^i}{\approxparamvec_{r}^i}{r}\Kernelsub{\approxparamvec_{r}^i}{\paramvec_{r-1}^i}{r}} \notag\\
	&= \frac{\CondPDF{\paramvec_R^i}{\discrep{\dat}{\simdat} \leq \epsilon_R}}{\PDF{\paramvec_{0}^i}}\prod_{r=1}^R\frac{\approxBKernel{\approxparamvec_{r}^i}{\paramvec_{r}^i}{r}\BKernel{\paramvec_{r-1}^i}{\approxparamvec_{r}^i}{r-1}}{\approxKernelsub{\paramvec_{r}^i}{\approxparamvec_{r}^i}{r}\Kernelsub{\approxparamvec_{r}^i}{\paramvec_{r-1}^i}{r}}\notag \\
	&= \frac{\CondPDF{\paramvec_R^i}{\discrep{\dat}{\simdat} \leq \epsilon_R}}{\PDF{\paramvec_{0}^i}}\frac{\prod_{r=1}^R B_{r-1}(\paramvec_{r-1}^i \mid \paramvec_{r}^i)}{\prod_{r=1}^R F_r(\paramvec_{r}^i \mid \paramvec_{r-1}^i)}, 
	\label{eq:expw}
	\end{align}
	where $F_r(\paramvec_{r}^i \mid \paramvec_{r-1}^i) = \approxKernelsub{\paramvec_{r}^i}{\approxparamvec_{r}^i}{r}\Kernelsub{\approxparamvec_{r}^i}{\paramvec_{r-1}^i}{r}$ is the composite proposal kernel and $B_{r-1}(\paramvec_{r-1}^i \mid \paramvec_{r}^i) = \BKernel{\paramvec_{r-1}^i}{\approxparamvec_{r}^i}{r-1}\approxBKernel{\approxparamvec_{r}^i}{\paramvec_{r}^i}{r}$ is the composite backward kernel. We observe that \eqref{eq:expw} is equivalent to the weight obtained from direct importance sampling on the joint space of the entire particle trajectory~\cite{DelMoral2006,Sisson2007}, that is,
	\begin{equation*}
	w_R^i = \frac{\pi_R(\paramvec_{0}^i,\paramvec_{1}^i,\ldots,\paramvec_{R}^i)}{\pi_0(\paramvec_{0}^i,\paramvec_{1}^i,\ldots,\paramvec_{R}^i)}.
	\end{equation*}
	Here, the importance distribution, given by
	\begin{align*}
	\pi_0(\paramvec_{0}^i,\paramvec_{1}^i,\ldots,\paramvec_{R}^i)= \PDF{\paramvec_{0}^i}\prod_{r=1}^R F_r(\paramvec_{r}^i \mid \paramvec_{r-1}^i),
	\end{align*}
	is the process of sampling from the prior and performing a sequence of kernel transitions. Finally, we note that the target distribution admits the target ABC posterior as a marginal density, that is, 
	\begin{align*}
	\int_{\mathbb{R}^{R}}\pi_R(\paramvec_{0}^i,\paramvec_{1}^i,\ldots,\paramvec_{R}^i)\, \text{d}\paramvec_{0}^i\ldots\text{d}\paramvec_{R-1}^i = \CondPDF{\paramvec_R^i}{\discrep{\dat}{\simdat} \leq \epsilon_R}.
	\end{align*}
	Therefore, for any function $f(\cdot)$ that is integrable with respect to the ABC posterior measure we have,
	\begin{equation*}
	\sum_{i=1}^{\mathcal{M}} f(\paramvec_{R}^i)w_R^i \to \int_{\mathbb{R}} f(\paramvec_{R})\CondPDF{\paramvec_R}{\discrep{\dat}{\simdat} \leq \epsilon_R}\, \text{d}\paramvec_R =  \E{f(\paramvec_{R})},
	\end{equation*}
	as $\mathcal{M} \to \infty$, that is, the PC-SMC-ABC method is unbiased. 
	
	\section{Example of MM-SMC-ABC with the  Ornstein-Uhlenbeck process}
	The Ornstein-Uhlenbeck process~\cite{Ornstein1930} (Equation (11)) is used in the main manuscript (Section 3.1) to demonstrate the PC-SMC-ABC. Here, we replicate the demonstration for MM-SMC-ABC. 
	
	Following the main manuscript, we use Euler-Maruyama simulations~\cite{Maruyama1955} with time discretisation $\Delta t = 0.01$ (Equation (13) ) for the exact model, and the stationary distribution of the Ornstein-Uhlenbeck process (Equation (14)) for the approximate model. Similarly, the main manuscript is followed for the data $\mathcal{D} = [X_T^1,X_T^2,\ldots, X_T^N]$ which is generated with $N = 1000$, $T = 1$, $x_0 = 10$, $\mu = 1$, $\gamma = 2$ , and $\sigma = 2\sqrt{5}$. 
	To better visually demonstrate MM-SMC-ABC, we infer the joint distribution of $\paramvec = (\mu, D)$ where $D= \sigma^2/2$ using MM-ABC-SMC (Algorithm 3 in main manuscript) with $\mathcal{M}$ particles, tuning parameter $\alpha = 0.1$, and ABC threshold sequence $\epsilon_r = \epsilon_{r-1}/2$ for $r = 1,2,\ldots 7$ with $\epsilon_{0}= 6.4$. 
	
\begin{figure}[h]
		\centering
		\includegraphics[width=\textwidth]{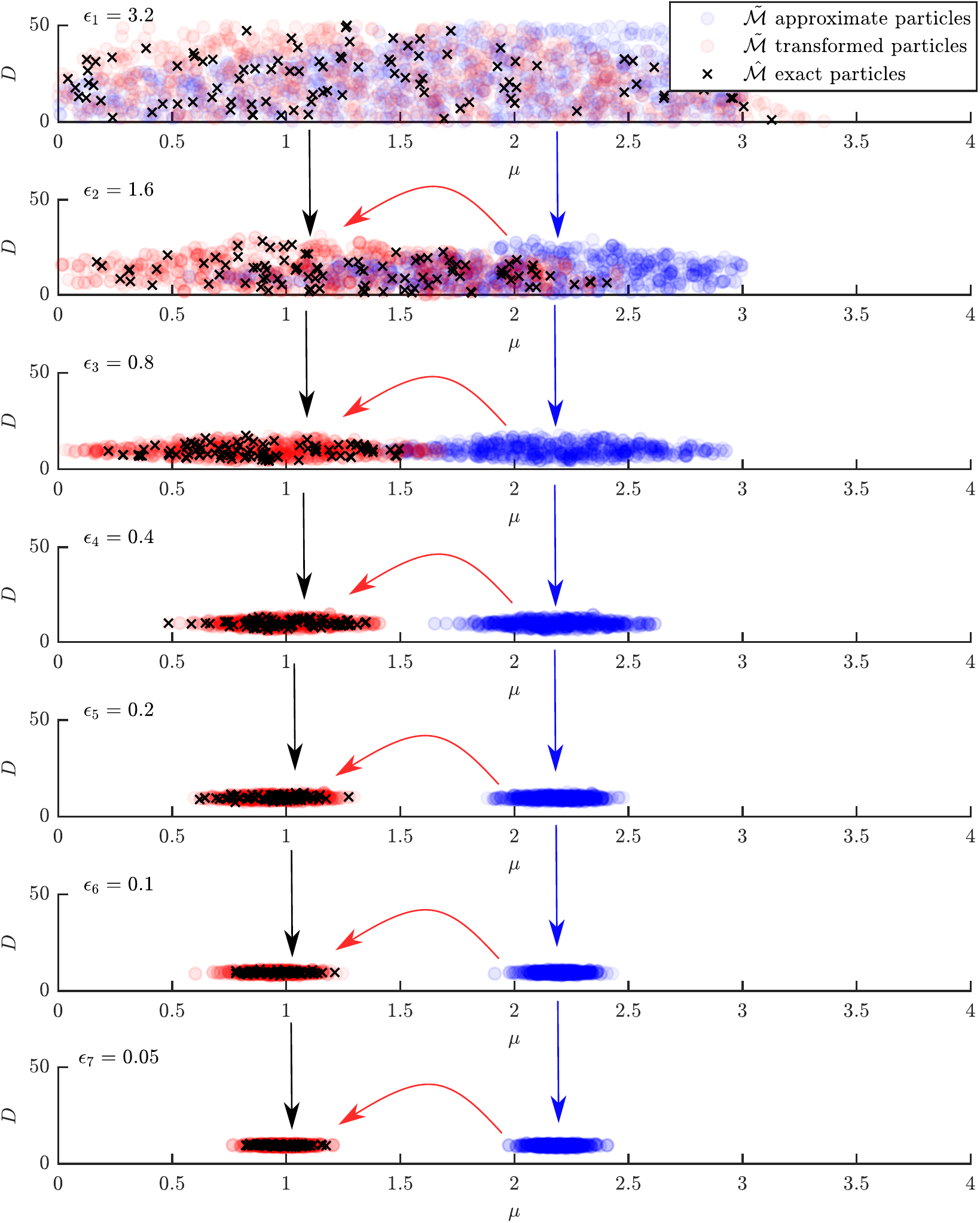}
		\caption{MM-SMC-ABC intermediate steps for the Ornstein-Uhlenbeck SDE example. Each panel demonstrates the SMC-ABC particles using the approximate model (blue dots), the small subset of exact particles (black crosses), and the transformed approximate particles (red dots) to form a new proposal distribution for the MM-SMC-ABC sampler.}
		\label{fig:mm-abc-smc-ou}
	\end{figure}

	Figure~\ref{fig:mm-abc-smc-ou} demonstrates the movement of  $\tilde{\mathcal{M}}= \floor{(1-\alpha)\mathcal{M}}$ approximate particles (blue dots), that are transformed (red dots) using $\hat{\mathcal{M}} = \ceil{\alpha\mathcal{M}}$ exact particles (black crosses) according to Equation (10) from the main manuscript. For this example, the speedup factor is approximately $10\times$.  
	
	\FloatBarrier
	\newpage
	
	\section{Derivation of approximate continuum-limit description}
	
	Here we derive the approximate continuum-limit description for our hexagonal lattice based discrete random walk model with generalized crowding function $f : [0,1] \to [-1,1]$. We follow the method of Simpson~et al.,~\cite{Simpson2010} and Jin et~al.,~\cite{Jin2016b}.
	
	Our main modification is dealing with a potentially negative crowding function. To deal with this case, we define two auxiliary functions,
	\begin{equation}
	f^+(C) = \begin{cases}
	f(C) & \text{if } f(C) \geq 0,\\
	0 & \text{otherwise},
	\end{cases}, \quad f^-(C) = \begin{cases}
	|f(C)| & \text{if } f(C) < 0,\\
	0 & \text{otherwise}.
	\end{cases}
	\label{eq:fcomp}
	\end{equation} 
	Note that $f(C) = f^+(C) - f^-(C)$, as this is important later.
	
	We assume a mean-field, that is, for any two lattice sites, $(x_1,y_1),(x_2,y_2)\in \mathbb{R}^2$, then their occupancy probabilities are independent, which results in the property,\\ $\E{C((x_1,y_1),t)C((x_2,y_2),t)} = \E{C((x_1,y_1),t)}\E{C((x_2,y_2),t)}$. Using this property, we denote $\mathcal{C}(x,y,t) = \E{C((x,y),t)}$ and write the conservation of probability equation that describes the change in occupancy probability of a site over a single time step,
	\begin{align}
	\begin{split}
	\Delta\mathcal{C}(x,y,t)=&\, P_m\left(1- \mathcal{C}(x,y,t)\right)\hat{\mathcal{C}}(x,y,t) - P_m\mathcal{C}(x,y,t)\left(1 - \hat{\mathcal{C}}(x,y,t)\right)  \\
	&+ \frac{P_p}{6}\left(1-\mathcal{C}(x,y,t)\right)\left(\sum_{(x',y')\in N(x,y)}\mathcal{C}(x',y',t)\frac{f^+(\hat{\mathcal{C}}(x',y',t)}{1-\hat{\mathcal{C}}(x',y',t)}\right)  \\
	&- P_p \mathcal{C}(x,y,t)f^-(\hat{\mathcal{C}}(x,y,t),
	\end{split}
	\label{eq:conserv}
	\end{align}
	where $\Delta\mathcal{C}(x,y,t) = \mathcal{C}(x,y,t+\tau) - \mathcal{C}(x,y,t)$,
	\begin{equation}
	\hat{\mathcal{C}}(x,y,t) = \frac{1}{6}\sum_{(x',y')\in N(x,y)}\mathcal{C}(x',y',t),
	\end{equation}
	and
	\begin{equation}
	\hat{\mathcal{C}}(x',y',t) = \frac{1}{6}\sum_{(x'',y'')\in N(x',y')}\mathcal{C}(x'',y'',t).
	\end{equation}
	The conservation of probability equation (Equation~\ref{eq:conserv}) deserves some interpretation. The first term is the probability that site $(x,y)$ is unoccupied at time $t$ and becomes occupied at time $t+\tau$ due to a successful motility event that moves an agent from an occupied neighboring site into $(x,y)$. The second term is the probability that site $(x,y)$ is occupied at time $t$ and becomes unoccupied at time $t+\tau$	due to a successful motility event that moves the agent away to a neighboring site. The third term is the probability that site $(x,y)$ is unoccupied at time $t$ and becomes occupied at time $t + \tau$ due to a successful proliferation event from an occupied neighboring site. The final term is the probability that the site $(x,y)$ is occupied at time $t$ and becomes unoccupied at time $t + \tau$ due to $f(\hat{C}((x,y),t)) < 0$.
	
\begin{figure}
		\centering
		\includegraphics[width=0.5\linewidth]{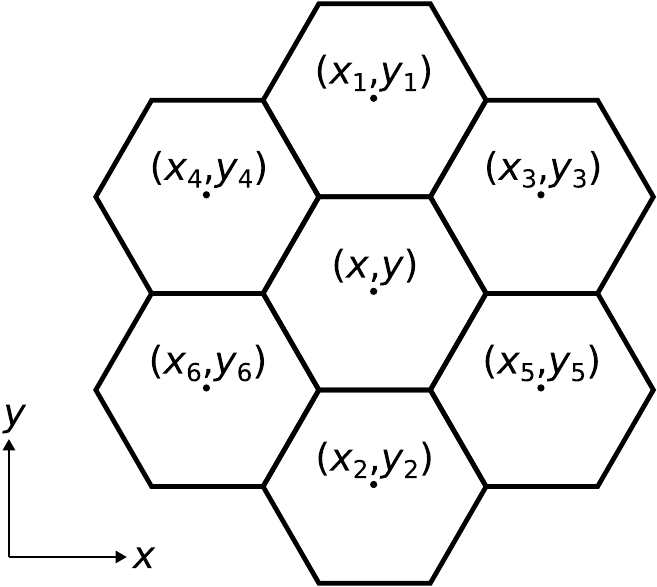}
		\caption{Schematic of the local neighborhood for the hexagonal lattice. Neighbor coordinates are $(x_1,y_1) = (x,y + \delta)$, $(x_2,y_2) = (x,y - \delta)$, $(x_3,y_3) = (x+ \delta\sqrt{3}/2,y + \delta/2)$,\\ $(x_4,y_4) = (x - \delta\sqrt{3}/2,y + \delta/2)$, $(x_5,y_5) = (x+ \delta\sqrt{3}/2,y - \delta/2)$, and \\$(x_6,y_6) = (x - \delta\sqrt{3}/2,y - \delta/2)$.}
		\label{fig:neighbour}
	\end{figure}

	For a hexagonal lattice site, $(x,y)$, we have six immediate neighboring lattice sites, $(x_1,y_1) = (x,y + \delta)$, $(x_2,y_2) = (x,y - \delta)$, $(x_3,y_3) = (x+ \delta\sqrt{3}/2,y + \delta/2)$,\\ $(x_4,y_4) = (x - \delta\sqrt{3}/2,y + \delta/2)$, $(x_5,y_5) = (x+ \delta\sqrt{3}/2,y - \delta/2)$, and \\$(x_6,y_6) = (x - \delta\sqrt{3}/2,y - \delta/2)$. The positions of these neighbors are shown in the schematic in Figure~ \ref{fig:neighbour}.
	We obtain an expression for $\mathcal{C}(x,y,t)$ at each of the six neighboring lattice points  by associating $\mathcal{C}(x,y,t)$ with a continuous function and performing a Taylor expansion about the point $(x,y)$.
	The result is,  
	\begin{align}
	\begin{split}	
	\mathcal{C}(x_1,y_1,t) =&\, \mathcal{C}(x,y,t) + \delta \pdydx{\mathcal{C}(x,y,t)}{y}  + \frac{\delta^2}{2}\pddydx{\mathcal{C}(x,y,t)}{y} +\mathcal{O}(\delta^3), \\
	\mathcal{C}(x_2,y_2,t) =&\, \mathcal{C}(x,y,t) - \delta \pdydx{\mathcal{C}(x,y,t)}{y}  + \frac{\delta^2}{2}\pddydx{\mathcal{C}(x,y,t)}{y} +\mathcal{O}(\delta^3), \\
	\mathcal{C}(x_3,y_3,t) =&\, \mathcal{C}(x,y,t) + \frac{\delta\sqrt{3}}{2} \pdydx{\mathcal{C}(x,y,t)}{x} + \frac{\delta}{2}\pdydx{\mathcal{C}(x,y,t)}{y}  \\&+ \frac{\delta^2}{2}\left[  \frac{3}{4}\pddydx{\mathcal{C}(x,y,t)}{x} + \frac{\sqrt{3}}{2}\pmddydx{\mathcal{C}(x,y,t)}{x}{y} + \frac{1}{4}\pddydx{\mathcal{C}(x,y,t)}{y}\right] +\mathcal{O}(\delta^3), \\
	\mathcal{C}(x_4,y_4,t) =&\, \mathcal{C}(x,y,t) - \frac{\delta\sqrt{3}}{2} \pdydx{\mathcal{C}(x,y,t)}{x} + \frac{\delta}{2}\pdydx{\mathcal{C}(x,y,t)}{y}  \\&+ \frac{\delta^2}{2}\left[  \frac{3}{4}\pddydx{\mathcal{C}(x,y,t)}{x} - \frac{\sqrt{3}}{2}\pmddydx{\mathcal{C}(x,y,t)}{x}{y} + \frac{1}{4}\pddydx{\mathcal{C}(x,y,t)}{y}\right] +\mathcal{O}(\delta^3), \\
	\mathcal{C}(x_5,y_5,t) =&\, \mathcal{C}(x,y,t) + \frac{\delta\sqrt{3}}{2} \pdydx{\mathcal{C}(x,y,t)}{x} - \frac{\delta}{2}\pdydx{\mathcal{C}(x,y,t)}{y}  \\&+ \frac{\delta^2}{2}\left[  \frac{3}{4}\pddydx{\mathcal{C}(x,y,t)}{x} + \frac{\sqrt{3}}{2}\pmddydx{\mathcal{C}(x,y,t)}{x}{y} + \frac{1}{4}\pddydx{\mathcal{C}(x,y,t)}{y}\right] +\mathcal{O}(\delta^3), \\
	\mathcal{C}(x_6,y_6,t) =&\, \mathcal{C}(x,y,t) - \frac{\delta\sqrt{3}}{2} \pdydx{\mathcal{C}(x,y,t)}{x} - \frac{\delta}{2}\pdydx{\mathcal{C}(x,y,t)}{y}  \\&+ \frac{\delta^2}{2}\left[  \frac{3}{4}\pddydx{\mathcal{C}(x,y,t)}{x} + \frac{\sqrt{3}}{2}\pmddydx{\mathcal{C}(x,y,t)}{x}{y} + \frac{1}{4}\pddydx{\mathcal{C}(x,y,t)}{y}\right] +\mathcal{O}(\delta^3). 
	\end{split}
	\label{eq:taylor_exp}
	\end{align}
	Therefore, we obtain an expression for $\hat{\mathcal{C}}(x,y,t)$,
	\begin{align}
	\hat{\mathcal{C}}(x,y,t) &= \frac{1}{6}\sum_{(x',y')\in N(x,y)}\mathcal{C}(x',y',t) \notag\\&= \frac{1}{6}\sum_{k=1}^6\mathcal{C}(x_k,y_k,t) \notag\\
	&= \mathcal{C}(x,y,t) + \frac{\delta^2}{4}\left[\pddydx{\mathcal{C}(x,y,t)}{x}+\pddydx{\mathcal{C}(x,y,t)}{y}\right] +\mathcal{O}(\delta^3).
	\label{eq:localsum}
	\end{align}
	This expression is required for the first, second, and fourth terms in the conservation equation (\eqref{eq:conserv}).
	
	To deal with the third term in \eqref{eq:conserv} we require an expression for\\ $g^+(\hat{\mathcal{C}}(x_k,y_k,t)) = f^+(\hat{\mathcal{C}}(x_k,y_k,t))/(1-\hat{\mathcal{C}}(x_k,y_k,t))$, $k = 1,2,\ldots,6$. By combining \eqref{eq:taylor_exp} and \eqref{eq:localsum} we obtain
	\begin{align}
	\begin{split}
	\hat{\mathcal{C}}(x_1,y_1,t) =&\, \mathcal{C}(x,y,t) + \delta \pdydx{\mathcal{C}(x,y,t)}{y} + \frac{\delta^2}{4}\left[\pddydx{\mathcal{C}(x,y,t)}{x}+3\pddydx{\mathcal{C}(x,y,t)}{y}\right]+\mathcal{O}(\delta^3)\\
	\hat{\mathcal{C}}(x_2,y_2,t) =&\, \mathcal{C}(x,y,t) + \delta \pdydx{\mathcal{C}(x,y,t)}{y} + \frac{\delta^2}{4}\left[\pddydx{\mathcal{C}(x,y,t)}{x}+3\pddydx{\mathcal{C}(x,y,t)}{y}\right]+\mathcal{O}(\delta^3)\\
	\hat{\mathcal{C}}(x_3,y_3,t) =&\, \mathcal{C}(x,y,t) + \frac{\delta\sqrt{3}}{2} \pdydx{\mathcal{C}(x,y,t)}{x} + \frac{\delta}{2} \pdydx{\mathcal{C}(x,y,t)}{y} \\& + \frac{\delta^2}{4}\left[\frac{5}{2}\pddydx{\mathcal{C}(x,y,t)}{x}+ \sqrt{3}\pddydx{\mathcal{C}(x,y,t)}{x}{y}+\frac{3}{2}\pddydx{\mathcal{C}(x,y,t)}{y}\right]+\mathcal{O}(\delta^3)\\
	\hat{\mathcal{C}}(x_4,y_4,t) =&\, \mathcal{C}(x,y,t) - \frac{\delta\sqrt{3}}{2} \pdydx{\mathcal{C}(x,y,t)}{x} + \frac{\delta}{2} \pdydx{\mathcal{C}(x,y,t)}{y} \\& + \frac{\delta^2}{4}\left[\frac{5}{2}\pddydx{\mathcal{C}(x,y,t)}{x}- \sqrt{3}\pddydx{\mathcal{C}(x,y,t)}{x}{y}+\frac{3}{2}\pddydx{\mathcal{C}(x,y,t)}{y}\right]+\mathcal{O}(\delta^3),\\
	\hat{\mathcal{C}}(x_5,y_5,t) =&\, \mathcal{C}(x,y,t) + \frac{\delta\sqrt{3}}{2} \pdydx{\mathcal{C}(x,y,t)}{x} - \frac{\delta}{2} \pdydx{\mathcal{C}(x,y,t)}{y} \\& + \frac{\delta^2}{4}\left[\frac{5}{2}\pddydx{\mathcal{C}(x,y,t)}{x} - \sqrt{3}\pddydx{\mathcal{C}(x,y,t)}{x}{y}+\frac{3}{2}\pddydx{\mathcal{C}(x,y,t)}{y}\right]+\mathcal{O}(\delta^3),\\
	\hat{\mathcal{C}}(x_3,y_3,t) =&\, \mathcal{C}(x,y,t) - \frac{\delta\sqrt{3}}{2} \pdydx{\mathcal{C}(x,y,t)}{x} - \frac{\delta}{2} \pdydx{\mathcal{C}(x,y,t)}{y} \\& + \frac{\delta^2}{4}\left[\frac{5}{2}\pddydx{\mathcal{C}(x,y,t)}{x}+ \sqrt{3}\pddydx{\mathcal{C}(x,y,t)}{x}{y}+\frac{3}{2}\pddydx{\mathcal{C}(x,y,t)}{y}\right]+\mathcal{O}(\delta^3).
	\end{split}
	\label{eq:taylor_localdense}
	\end{align}
	Each expression in \eqref{eq:taylor_localdense} is of the form $\hat{\mathcal{C}}(x_k,y_k,t) = \mathcal{C}(x,y,t) + \bar{\mathcal{C}}_k$, where\\ $\bar{\mathcal{C}}_k = \mathcal{O}(\delta)$. This allows us to consider the Taylor expansion of $g^+(\hat{\mathcal{C}}(x_k,y_k,t))$ about the density $\mathcal{C}(x,y,t)$, that is,
	\begin{equation}
	g^+(\mathcal{C} + \bar{\mathcal{C}}_k) = g^+(\mathcal{C}) + \bar{\mathcal{C}}_k\dydx{g^+(\mathcal{C})}{\mathcal{C}} + \frac{\bar{\mathcal{C}}_k^2}{2}\ddydx{g^+(\mathcal{C})}{\mathcal{C}} + \mathcal{O}(\delta^3).
	\label{eq:taylor_g}
	\end{equation}
	Using \eqref{eq:taylor_g}, we can obtain an expression for the summation within the third term of the conservation equation (\eqref{eq:conserv}), that is,
	\begin{align}
	\sum_{(x',y')\in N(x,y)}\mathcal{C}(x',y',t)\frac{f^+(\hat{\mathcal{C}}(x',y',t)}{1-\hat{\mathcal{C}}(x',y',t)} =&\, \sum_{k=1}^6\mathcal{C}(x_k,y_k,t)g^+(\mathcal{C}(x,y,t)+ \bar{\mathcal{C}}_k) \notag\\
	=&\, g^+(\mathcal{C}(x,y,t)) \sum_{k=1}^6\mathcal{C}(x_k,y_k,t) + \dydx{g^+(\mathcal{C})}{\mathcal{C}}\sum_{k=1}^6 \bar{\mathcal{C}}_k\mathcal{C}(x_k,y_k,t) \notag \\ &+ \ddydx{g^+(\mathcal{C})}{\mathcal{C}}\sum_{k=1}^6 \frac{\bar{\mathcal{C}}_k^2}{2}\mathcal{C}(x_k,y_k,t) + \mathcal{O}(\delta^3).
	\label{eq:taylor_prolif}
	\end{align}
	After substitution of \eqref{eq:taylor_localdense} into \eqref{eq:taylor_prolif} and some tedious algebra we arrive at
	\begin{align}
	\sum_{(x',y')\in N(x,y)}\mathcal{C}(x',y',t)\frac{f^+(\hat{\mathcal{C}}(x',y',t)}{1-\hat{\mathcal{C}}(x',y',t)} =&\, 6\mathcal{C}(x,y,t)g^+(\mathcal{C}(x,y,t)) \notag\\ &+ \frac{3\delta^2}{2} g^+(\mathcal{C}(x,y,t))\left[\pddydx{\mathcal{C}(x,y,t)}{x}+\pddydx{\mathcal{C}(x,y,t)}{y}\right] \notag\\ &+ 3\delta^2\mathcal{C}(x,y,t)\dydx{g^+(\mathcal{C})}{\mathcal{C}}\left[\pddydx{\mathcal{C}(x,y,t)}{x}+\pddydx{\mathcal{C}(x,y,t)}{y}\right] \notag\\ &+ 3\delta^2\dydx{g^+(\mathcal{C})}{\mathcal{C}}\left[\left(\pdydx{\mathcal{C}(x,y,t)}{x}\right)^2+\left(\pdydx{\mathcal{C}(x,y,t)}{y}\right)^2\right] \notag \\ &+
	\frac{3\delta^2}{2}\mathcal{C}(x,y,t)\ddydx{g^+(\mathcal{C})}{\mathcal{C}}\left[\left(\pdydx{\mathcal{C}(x,y,t)}{x}\right)^2+\left(\pdydx{\mathcal{C}(x,y,t)}{y}\right)^2\right] \notag\\&+ \mathcal{O}(\delta^3).
	\label{eq:third_term}
	\end{align}
	Finally, we substitute \eqref{eq:localsum} and \eqref{eq:third_term} into \eqref{eq:conserv} to obtain
	\begin{align}
	\Delta\mathcal{C}(x,y,t)=& P_m\left(1- \mathcal{C}(x,y,t)\right)\left[\mathcal{C}(x,y,t) + \frac{\delta^2}{4} \nabla^2\mathcal{C}(x,y,t)\right]\notag\\ &- P_m\mathcal{C}(x,y,t)\left[1 - \mathcal{C}(x,y,t) -\frac{\delta^2}{4} \nabla^2\mathcal{C}(x,y,t) \right]\notag  \\
	&+ \frac{P_p}{6}\left(1-\mathcal{C}(x,y,t)\right)\left[
	6\mathcal{C}(x,y,t)g^+(\mathcal{C}(x,y,t)) + \frac{3\delta^2}{2} g^+(\mathcal{C}(x,y,t))\nabla^2\mathcal{C}(x,y,t) \right.\notag\\ &+ 3\delta^2\mathcal{C}(x,y,t)\dydx{g^+(\mathcal{C})}{\mathcal{C}}\nabla^2\mathcal{C}(x,y,t)+ 3\delta^2\dydx{g^+(\mathcal{C})}{\mathcal{C}}\left(\nabla\mathcal{C}(x,y,t)\cdot\nabla\mathcal{C}(x,y,t)\right) \notag\\ &+
	\left.\frac{3\delta^2}{2}\mathcal{C}(x,y,t)\ddydx{g^+(\mathcal{C})}{\mathcal{C}}\left(\nabla\mathcal{C}(x,y,t)\cdot\nabla\mathcal{C}(x,y,t)\right) \right] \notag \\
	&- P_p \mathcal{C}(x,y,t)f^-(\mathcal{C}(x,y,t) + \mathcal{O}(\delta^3).
	\label{eq:conserv2}
	\end{align}	
	Here we have used the notation $\nabla^2\mathcal{C} = \pddydx{\mathcal{C}}{x} + \pddydx{\mathcal{C}}{y}$, and $\nabla\mathcal{C \cdot\nabla \mathcal{C}} = (\pdydx{\mathcal{C}}{x})^2 + (\pdydx{\mathcal{C}}{y})^2$. After rearranging \eqref{eq:conserv2}, we obtain
	\begin{align}
	\Delta\mathcal{C}(x,y,t)=& \frac{P_m\delta^2}{4}\nabla^2\mathcal{C}(x,y,t)
	+ P_p\mathcal{C}(x,y,t)\left[\left(1-\mathcal{C}(x,y,t)\right)g^+(\mathcal{C}(x,y,t)) - f^-(\mathcal{C}(x,y,t))\right] \notag \\
	&+ P_p \delta^2 H(\mathcal{C}(x,y,t)) + \mathcal{O}(\delta^3)
	\label{eq:conserv3}
	\end{align}
	where 
	\begin{align*}
	H(\mathcal{C}(x,y,t)) =& 
	\frac{1}{4} g^+(\mathcal{C}(x,y,t))\nabla^2\mathcal{C}(x,y,t) + \frac{1}{2}\mathcal{C}(x,y,t)\dydx{g^+(\mathcal{C})}{\mathcal{C}}\nabla^2\mathcal{C}(x,y,t)\notag \\&+  \frac{1}{2}\dydx{g^+(\mathcal{C})}{\mathcal{C}}\left(\nabla\mathcal{C}(x,y,t)\cdot\nabla\mathcal{C}(x,y,t)\right) +\frac{1}{4}\mathcal{C}(x,y,t)\ddydx{g^+(\mathcal{C})}{\mathcal{C}}\left(\nabla\mathcal{C}(x,y,t)\cdot\nabla\mathcal{C}(x,y,t)\right). 
	\end{align*}
	Note that $f^+(\mathcal{C}(x,y,t)) = g^+(\mathcal{C}(x,y,t))(1 - \mathcal{C}(x,y,t))$ and, by \eqref{eq:fcomp}, that\\ $f(\mathcal{C}(x,y,t)) = f^+(\mathcal{C}(x,y,t)) - f^-(\mathcal{C}(x,y,t))$. Then \eqref{eq:conserv3} becomes
	\begin{align*}
	\Delta\mathcal{C}(x,y,t)= \frac{P_m\delta^2}{4}\nabla^2\mathcal{C}(x,y,t)
	+ P_p\mathcal{C}(x,y,t)f(\mathcal{C}(x,y,t)) + P_p \delta^2 H(\mathcal{C}(x,y,t)) + \mathcal{O}(\delta^3).
	%\label{eq:conserv4}
	\end{align*}
	After dividing by the time interval, $\tau$, and choosing $\delta^2 = \mathcal{O}(\tau)$ and $P_p = \mathcal{O}(\tau)$, then we have the following limits,
	\begin{align*}
	\lim_{\tau \to 0} \frac{P_m \delta^2}{4 \tau} = D, \quad \lim_{\tau \to 0} \frac{P_p}{\tau} = \lambda, \quad\lim_{\tau \to 0} \frac{\Delta\mathcal{C}(x,y,t)}{\tau} = \pdydx{\mathcal{C}(x,y,t)}{t}, \quad
	\lim_{\tau \to 0} \frac{P_p\delta^2}{\tau} = 0.
	\end{align*}
	Therefore, we arrive at the continuum-limit approximation
	\begin{align*}
	\pdydx{\mathcal{C}(x,y,t)}{t}= D\nabla^2\mathcal{C}(x,y,t)
	+ \lambda\mathcal{C}(x,y,t)f(\mathcal{C}(x,y,t)).
	%\label{eq:conserv4}
	\end{align*}
	
	\newpage
	\FloatBarrier
	
	\section{Forward problem simulation}
	%\label{sec:methods}
	This section presents standard computational methods that are applied to \emph{forwards problems} of the models we consider in this work.
	
	\subsection{Stochastic simulation of the discrete model}
	In the work of Jin et al.~\cite{Jin2016b}, they consider crowding functions with $f : [0,1] \rightarrow [0,1]$. This is not quite sufficient to enable a crowding function with a carrying capacity $K < 1$ since motility events will cause the carrying capacity to be violated. Therefore, we take $f : [0,1] \rightarrow [-1,1]$. If $f(\hat{C}(\ell_s)) < 0$ then the site $\ell_s$ is removed from the set of occupied sites at time $t$, denoted by $\mathcal{L}(t)$, with probability $P_p|f(\hat{C}(\ell_s))|$.
	Given these definitions, the lattice-based random walk proceeds according to Algorithm~\ref{alg:lrw}.
	\begin{algorithm}[h]
		\caption{Lattice-based random walk model}
		\begin{algorithmic}[1]
			\State{Initialize $\mathcal{L}(0) \subset L$ with $|\mathcal{L}(0)| = N$ and $t \leftarrow 0$};
			\While{$t < T$}
			\State{$N \leftarrow |\mathcal{L}(t)|$};
			\For{$i = [1,2,\ldots, N]$}
			\State{Choose $\ell_s \in \mathcal{L}(t)$ uniformly at random with probability $1/{N}$};
			\State{Choose $\ell_m \in \mathcal{N}(\ell_s)$ uniformly at random with probability $1/{|\mathcal{N}(\ell_s)|}$};
			\If{$C(\ell_m,t) = 0$}
			\State{Generate $u \sim \mathcal{U}(0,1)$};
			\If{$u \leq P_m$}
			\State{Set $\mathcal{L}(t) \leftarrow \left\{\ell_m\right\} \cup \mathcal{L}(t) \backslash \left\{\ell_s\right\}$};
			\EndIf
			\EndIf 
			\EndFor
			\For{$i = [1,2,\ldots, N]$}
			\State{Choose $\ell_s \in \mathcal{L}(t)$ uniformly at random with probability $1/{N}$};
			\State{Set $\hat{C}(\ell_s) \leftarrow \left[1/{|\mathcal{N}(\ell_s)|}\right]\sum_{\ell_s' \in \mathcal{N}(\ell_s)} C(\ell_s',t)$};
			\State{Generate $u \sim \mathcal{U}(0,1)$};
			\If{$u \leq P_p|f(\hat{C}(\ell_s))|$}
			\If{$f(\hat{C}(\ell_s)) \geq 0$}
			\State{Choose $\ell_p \in \left\{ \ell_s' \in \mathcal{N}(\ell_s) : C(\ell_s',t) = 0\right\}$ uniformly at random with probability $1- \hat{C}(\ell_s)$};
			\State{Set $\mathcal{L}(t) \leftarrow \left\{\ell_p\right\} \cup \mathcal{L}(t)$};
			\Else
			\State{Set $\mathcal{L}(t) \leftarrow \mathcal{L}(t) \backslash \left\{\ell_s\right\}$};
			\EndIf
			\EndIf
			\EndFor
			\State{$t \leftarrow t + \tau$}
			\EndWhile
		\end{algorithmic}
		\label{alg:lrw}
	\end{algorithm}
	
	\FloatBarrier

	\subsection{Numerical solutions of continuum-limit differential equations}
	In our case, the approximate model is a deterministic continuum equation that is either a nonlinear ODE or PDE. In both cases we apply numerical schemes that automatically adapt the time step size to control the truncation error.
	
	\subsubsection{ODE numerical solutions}
	The Runge-Kutta-Fehlberg fourth-order-fifth-order method (RKF45)~\cite{Fehlberg1969} is a numerical scheme in the Runge-Kutta family of methods to approximate the solution of a nonlinear ODE of the form,
	\begin{equation*}
	\dydx{\mathcal{C}(t)}{t} = h(t,\mathcal{C}(t)), \quad 0 < t,
	\end{equation*}
	where $h(t, \mathcal{C}(t))$ is a function that satisfies certain regularity conditions and the initial condition, $\mathcal{C}(0)$.
	
	Given the approximate solution, $c_i \approx \mathcal{C}(t_i)$, an embedded pair of  Runge-Kutta methods, specifically a fourth and fifth order pair, are used to advanced the solution to $c_{i+1} \approx \mathcal{C}(t_i + \Delta t) + \mathcal{O}(\Delta t^4)$ and estimate the truncation error that can be used to adaptively adjust $\Delta t$ to ensure the error is always within some specified tolerance $\tau$.
	
	The fourth and fifth order estimates are, respectively,
	\begin{equation}
	c_{i+1} = c_{i} + \Delta t\left(\frac{25}{216}k_1 + \frac{1,408}{2,565}k_3 + \frac{2,197}{4,104}k_4 - \frac{1}{5}k_5\right),
	\label{eq:fourth}
	\end{equation}
	and
	\begin{equation}
	c^*_{i+1} = c_{i} + \Delta t \left(\frac{16}{135}k_1 + \frac{6,656}{12,825}k_3 + \frac{28,561}{56,430}k_4 - \frac{9}{50}k_5 + \frac{2}{55}k_6\right),
	\label{eq:fifth}
	\end{equation}
	where
	\begin{align*}
	k_1 &= h(t_i,c_i),\\
	k_2 &= h\left(t_i + \frac{\Delta t}{4}, c_i + \frac{\Delta t}{4}k_1 \right),\\
	k_3 &= h\left(t_i + \frac{3\Delta t}{8}, c_i + \Delta t\left[\frac{3}{32}k_1 + \frac{9}{32}k_2\right] \right),\\
	k_4 &= h\left(t_i + \frac{12\Delta t}{13}, c_i + \Delta t\left[\frac{1,932}{2,197}k_1 - \frac{7,200}{2,197}k_2 + \frac{7,296}{2,197}k_3\right] \right),\\
	k_5 &= h\left(t_i + \Delta t, c_i + \Delta t\left[\frac{439}{216}k_1 - 8 k_2 + \frac{3,680}{513}k_3 - \frac{845}{4,104}k_4\right] \right),\\
	k_6 &= h\left(t_i + \frac{\Delta t}{2}, c_i + \Delta t\left[-\frac{8}{27}k_1 + 2 k_2 - \frac{3,544}{2,565}k_3 + \frac{1,859}{4,104}k_4 - \frac{11}{40}k_5\right] \right).
	\end{align*}
	Note that the truncation error can be estimated by $\epsilon = |c_{i+1} - c^*_{i+1}|$. 
	After each evaluation of \eqref{eq:fourth} and \eqref{eq:fifth}, a new step size is determined by $\Delta t \leftarrow s\Delta t$, where $s = \left[\tau / (2\epsilon)\right]^{1/4}$. If $\epsilon \leq \tau$, then the solution is accepted and the new $\Delta t$ is used for the next iteration. Alternatively, if $\epsilon > \tau$ then the solution is rejected and a new attempt is made using the new time step. This process is repeated until the solution advances to some desired time point $T$. For more details and analysis of this method, see Iserles~\cite{Iserles2008}. 
	
	\subsubsection{PDE numerical solutions}
	In this work, we consider numerical solutions to PDEs of the form
	\begin{equation}
	\pdydx{\mathcal{C}(x,t)}{t} = D\pddydx{\mathcal{C}(x,t)}{x} + \lambda \mathcal{C}(x,t)f(\mathcal{C}(x,t)),\quad 0 < t, \quad 0 < x < L,
	\end{equation}
	with initial conditions
	\begin{equation*}
	\mathcal{C}(x,0) = c(x),\quad t = 0, 
	\end{equation*}
	and Neumann boundary conditions
	\begin{equation*}
	\pdydx{\mathcal{C}(x,t)}{x} = 0, \quad x = 0\text{ and }x = L.
	\end{equation*}
	
	Given the approximate solution, $c_{i}^j \approx \mathcal{C}(x_i,t_j)$ for $i = 1, 2, \ldots N$, then we apply a first order backward Euler discretization in time and first order central differences in space to yield
	\begin{align}
	\begin{split}
	\frac{c_2^{j+1}-c_1^{j+1}}{\Delta x } &= 0, \\
	\frac{c_i^{j+1} - c_i^j}{\Delta t} &= D\frac{c_{i+1}^{j+1} - 2c_i^{j+1} + c_{i-1}^{j+1}}{\Delta x^2} + \lambda c_i^{j+1}f(c_i^{j+1}), \quad i = 2,3, \ldots, N-1,\\
	\frac{c_N^{j+1}-c_{N-1}^{j+1}}{\Delta x } &= 0,
	\end{split}
	\label{eq:nonlin}
	\end{align}
	where $c_{i\pm1}^{j\pm1} \approx \mathcal{C}(x_i \pm \Delta x, t_j \pm \Delta t)$. The solution is stepped forward in time using fixed point iterations that are initialized by a first order forward Euler estimate.
	
	The truncation error is estimated by
	\begin{equation*}
	\epsilon = \frac{\Delta t}{2}\max_{1\leq i\leq N} \left|\left(\dydx{c}{t}\right)_i^{j+1} - \left(\dydx{c}{t}\right)_i^{j}\right|,\text{ with }
	\left(\dydx{c}{t}\right)_i^{j+1} \approx \frac{c_{i}^{j+1} -c_{i}^j }{\Delta t}.
	\end{equation*}
	After solving the nonlinear system (\eqref{eq:nonlin}) using fixed point iteration, a new step size is determined by $\Delta t \leftarrow s\Delta t$, where $s = 0.9\sqrt{\tau /\epsilon}$ and $\tau$ is the truncation error tolerance~\cite{Simpson2007,Sloan1999}. If $\epsilon \leq \tau$, then the solution is accepted and the new $\Delta t$ is used for the next time step. Alternatively, if $\epsilon > \tau$ then the solution is rejected and a new attempt is made using the new time step. This process is repeated until the solution advances to some desired time point $T$.

	\newpage
	\section{Additional results}
\begin{figure}[h]
		\centering
		\includegraphics[width=1\textwidth]{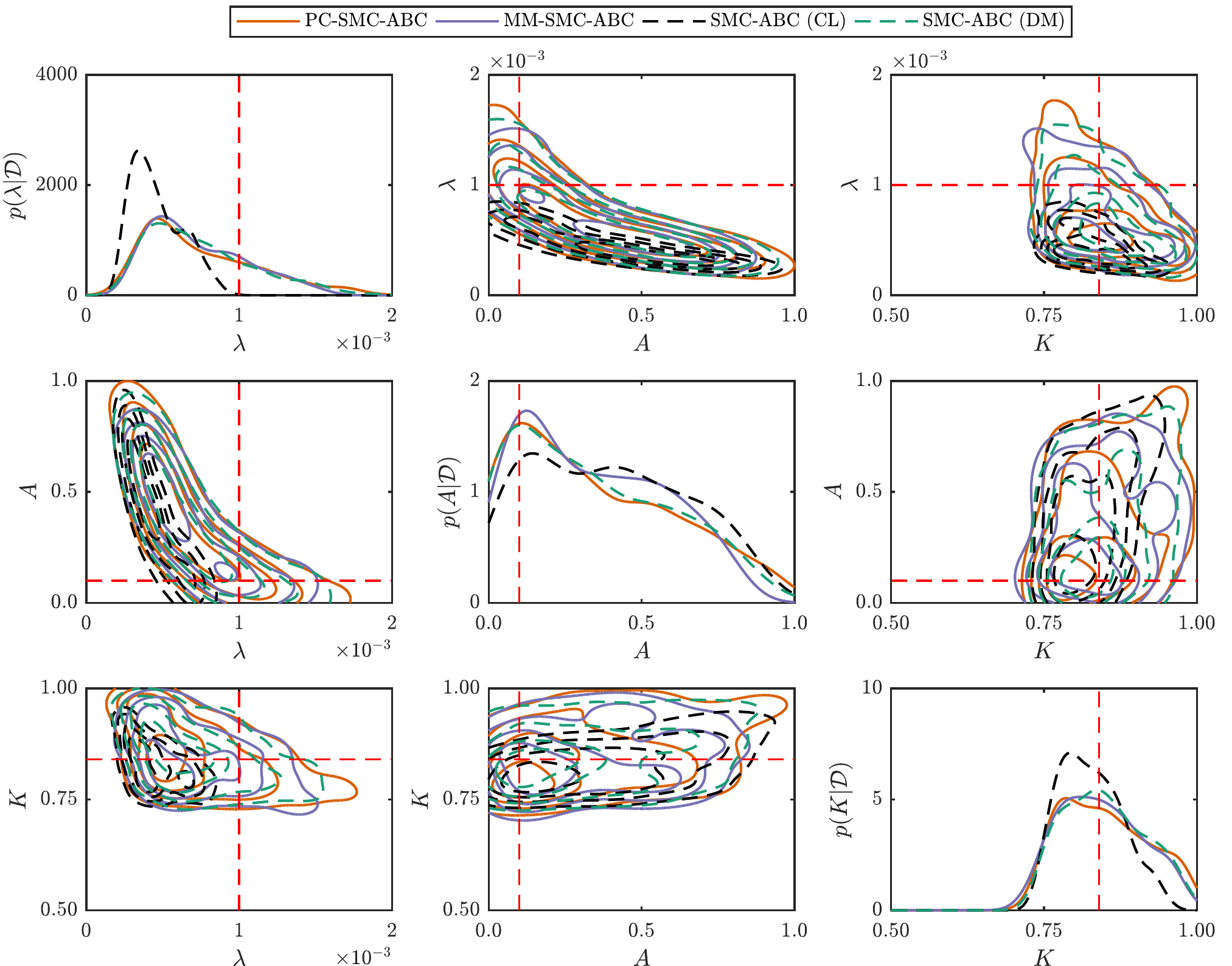}
		\caption{Comparison of estimated posterior marginal densities for the weak Allee model. There is a distinct bias in the SMC-ABC density estimate using the continuum limit (CL) (black dashed) compared with the SMC-ABC method with the discrete model (DM) (green dashed). However, the density estimates computed using the PC-SMC-ABC (orange solid) and MM-SMC-ABC (purple solid) methods match well with a reduced computational overhead.}
		\label{fig:alleemvcomp}
	\end{figure}

\begin{figure}
		\centering
		\includegraphics[width=1\linewidth]{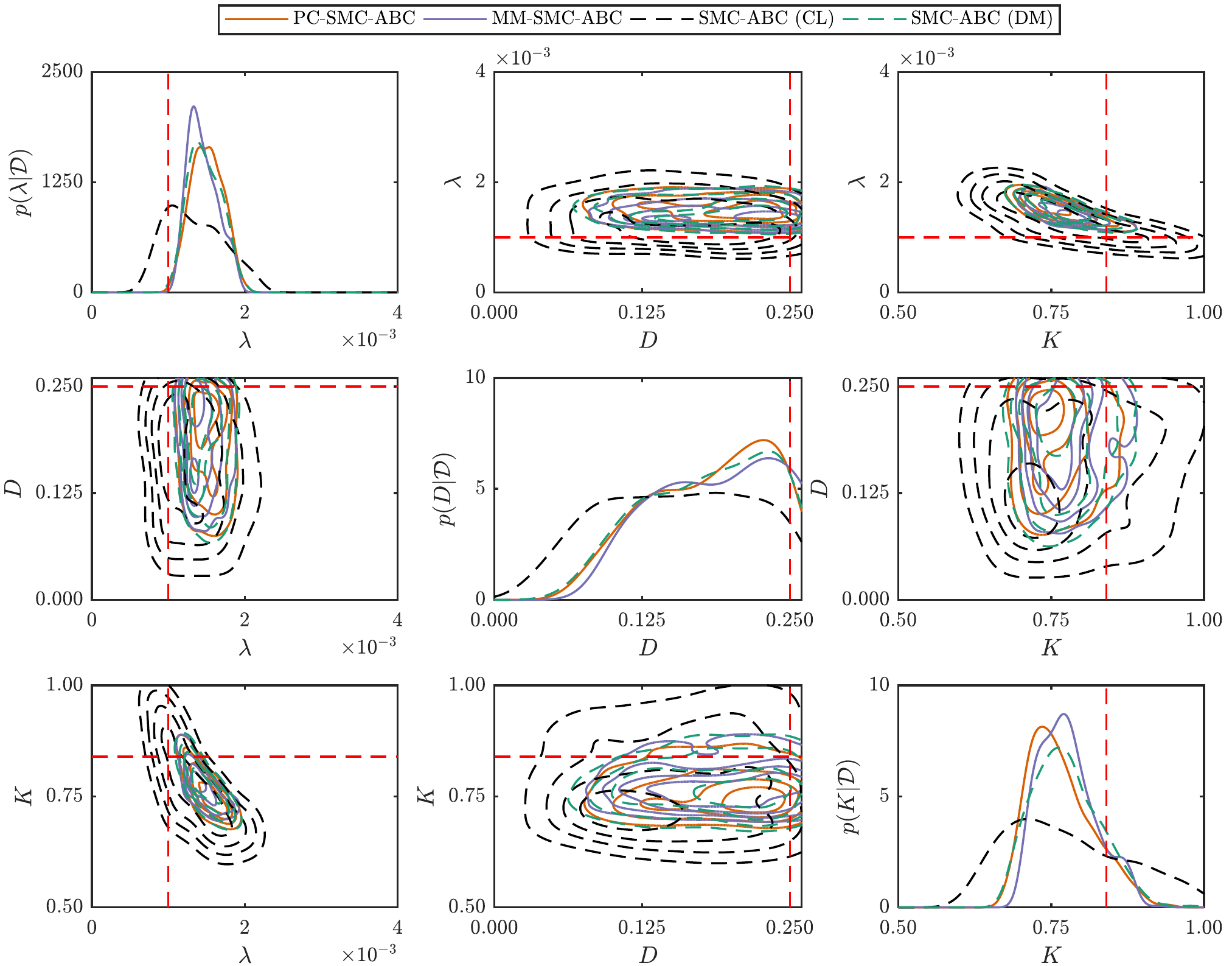}
		\caption{Comparison of estimated posterior marginal densities for the scratch assay model. There is a distinct bias in the SMC-ABC density estimate using the continuum limit (CL) (black dashed) compared with the SMC-ABC method with the discrete model (DM) (green dashed). However, the density estimates computed using the PC-SMC-ABC (orange solid) and MM-SMC-ABC (purple solid) methods match well with a reduced computational overhead.}
		\label{fig:fkppmvcomp}
	\end{figure}

	We provide the multivariate posteriors for the results provided in the main text. In both Figure~\ref{fig:alleemvcomp} and Figure~\ref{fig:fkppmvcomp} the contours of the posterior bivariate marginals generated, at a significant reduction in computational cost, both the PC-SMC-ABC and MM-SMC-ABC methods align very well with contours of the posterior bivariate marginals SMC-ABC using the expensive discrete model alone, however, the is a clear bias in the contours posterior bivariate marginals generated with SMC-ABC using the continuum-limit approximation alone. 
	
	\FloatBarrier
	\newpage
	\section{Effect of motility rate on continuum-limit approximation}
	
\begin{figure}[h]
		\centering
		\includegraphics[width=0.75\linewidth]{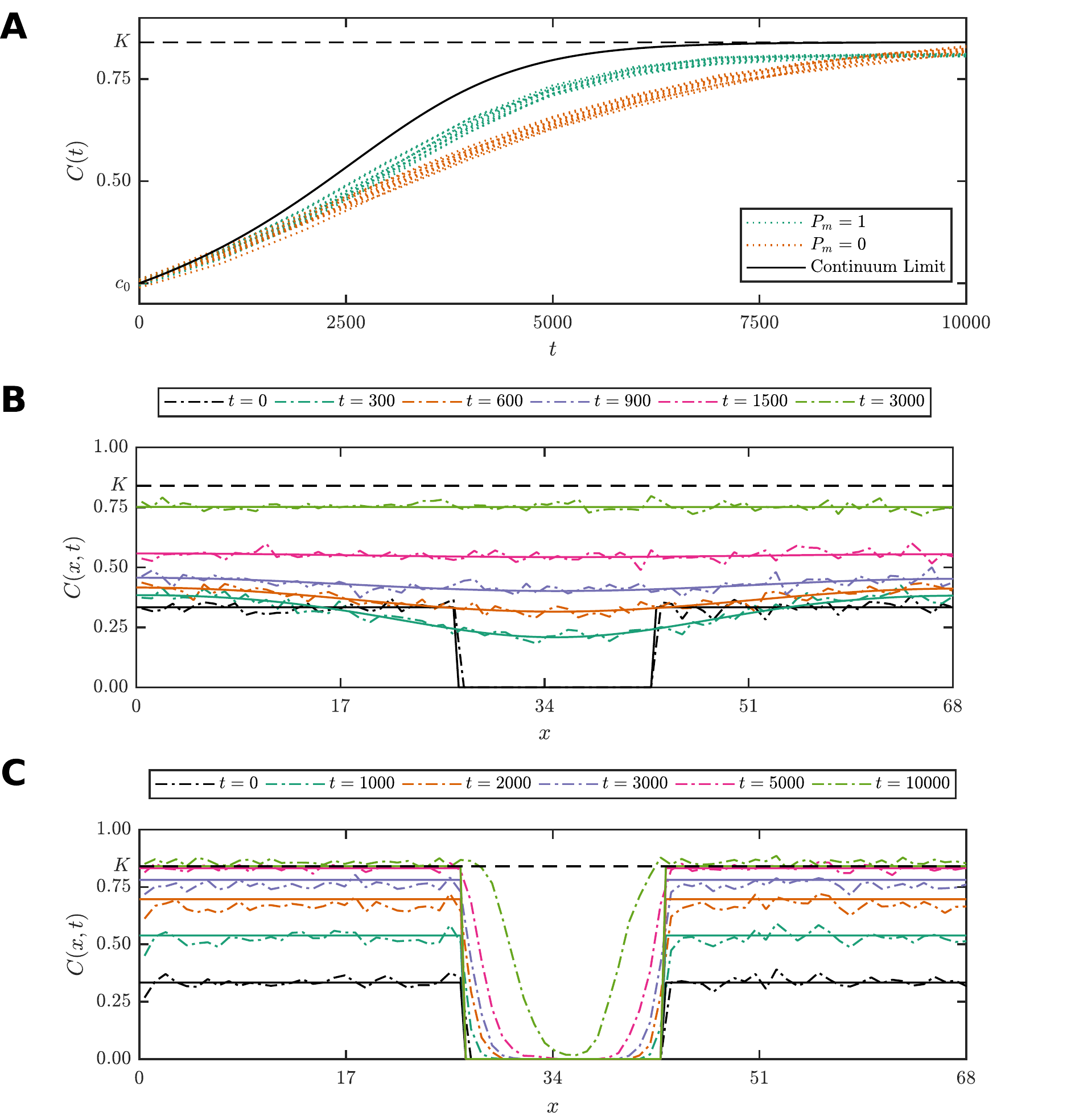}
		\caption{Continuum-limit approximations for the (A) weak Allee model and (B)-(C) the scratch assay model plotted against stochastic simulations of the discrete model. (A) For the weak Allee model, the continuum limit (black solid) does not capture the average behavior of the discrete model for motile agents, $P_m = 1$, (green dotted) or non-motile agents, $P_m = 0$, (orange dotted). (B)-(C) For the scratch assay model, the continuum limit (solid) is a very good match of the average behavior of the (B) discrete model (dot-dashed) for  motile agents, $P_m = 1$, but a very poor approximation of the average behavior of the (C) discrete model (dot-dashed) for non-motile agents, $P_m = 0$, especially in the scratch region. Parameters used in the simulations are $P_p = 1/1000$, $\lambda = P_p/\tau$ $D = P_m\delta^2/4\tau$, $K = 5/6$, and $A = 1/10$. The stochastic simulations are performed on an $I \times J$ hexagonal lattice with $I = 80$, $J = 68$, $\tau = 1$ and $\delta = 1$. } 
		\label{fig:clapprx}
	\end{figure}

	Here we demonstrate the ability (or lack thereof) of the continuum-limit approximation to capture the average behavior of the discrete model for the two examples considered in the main manuscript: the weak Allee model; and the scratch assay model. Figure~\ref{fig:clapprx} plots the solutions to the continuum-limit differential equation against realizations of the discrete model for both motile, $P_m = 1$, and non-motile, $P_m = 0$, agents.    
	
	In the case of the weak Allee model (Figure~\ref{fig:clapprx}(A)), the continuum limit does not match the average behavior in either case, though it does perform better when $P_m = 1$  than when $P_m = 0$. The remaining discrepancy when $P_m = 1$ is can be related to the ratio $P_p/P_m$ and to the effect of the neighborhood radius, $r$ (see Jin et~al.~\cite{Jin2016b} for further explanation). Decreasing $P_p/P_m$ and increasing $r$ will correct this discrepancy. The scratch assay continuum limit captures the average behavior of the discrete model very well when $P_m = 1$ (Figure~\ref{fig:clapprx}(B)), but does not capture the scratch closing behavior of the discrete model when $P_m = 0$ (Figure~\ref{fig:clapprx}(C)).
	
	\FloatBarrier
	
\end{appendices}

\end{document}